\documentclass[
aps,
prx,
twocolumn,
showpacs,
superscriptaddress
]{revtex4-2}

\usepackage{graphicx}
\usepackage{dcolumn}
\usepackage{bm}
\usepackage{amsmath,amssymb,mathtools}
\usepackage{color}
\usepackage{appendix}

\definecolor{darkblue}{rgb}{0,0,0.6}
\definecolor{darkred}{rgb}{0.6,0,0}
\definecolor{darkgreen}{rgb}{0,0.6,0}

\usepackage[colorlinks=true,urlcolor=darkblue,citecolor=darkblue,linkcolor=darkred,hyperfootnotes=false]{hyperref}

\usepackage{soul}
\usepackage{multirow}
\usepackage{tikz}
\usepackage{hhline}
\usepackage{algorithm}
\usepackage{algorithmic}

\usepackage[para,online,flushleft]{threeparttable}
\usepackage{booktabs}
\usepackage{multirow}

\usepackage{setspace}

\makeatother

\begin{document}


\title{Unified Hierarchical Relationship Between Thermodynamic Tradeoff Relations}

\author{Euijoon Kwon}
\affiliation{Department of Physics and Astronomy \& Center for Theoretical Physics, Seoul National University, Seoul 08826, Republic of Korea}
\affiliation{School of Physics, Korea Institute for Advanced Study, Seoul 02455, Republic of Korea}

\author{Jong-Min Park}
\affiliation{Asia Pacific Center for Theoretical Physics, Pohang, 37673, Republic of Korea}
\affiliation{Department of Physics, Postech, Pohang 37673, Republic of Korea}

\author{Jae Sung Lee}
\email{jslee@kias.re.kr}
\affiliation{School of Physics, Korea Institute for Advanced Study, Seoul 02455, Republic of Korea}

\author{Yongjoo Baek}
\email{y.baek@snu.ac.kr}
\affiliation{Department of Physics and Astronomy \& Center for Theoretical Physics, Seoul National University, Seoul 08826, Republic of Korea}

\date{\today}
 
\begin{abstract}
    Recent years have witnessed a surge of discoveries in the studies of thermodynamic inequalities: the thermodynamic uncertainty relation (TUR) and the entropic bound (EB) provide a lower bound on the entropy production (EP) in terms of nonequilibrium currents; the classical speed limit (CSL) expresses the lower bound on the EP using the geometry of probability distributions; the power-efficiency (PE) tradeoff dictates the maximum power achievable for a heat engine given the level of its thermal efficiency. In this study, we show that there exists a unified hierarchical structure encompassing all of these bounds, with the fundamental inequality given by an extension of the TUR (XTUR) that incorporates the most general range of current-like and state-dependent observables. By selecting more specific observables, the TUR and the EB follow from the XTUR, and the CSL and the PE tradeoff follow from the EB. Our derivations cover both Langevin and Markov jump systems, with the first proof of the EB for the Markov jump systems and a more generalized form of the CSL. We also present concrete examples of the EB for the Markov jump systems and the generalized CSL.
\end{abstract}


\maketitle

\section{Introduction}

The basis of thermodynamics lies in its fundamental inequalities. At their origin was the question of how much work a heat engine can do, which led to the discovery of the Carnot efficiency---a universal cap on the thermal efficiency for all heat engines~\cite{Carnot1824}. However, deeper explorations resulted in the discovery of an even more profound inequality, the second law of thermodynamics, which indicates the inevitable direction of evolution for all macroscopic thermodynamic systems~\cite{Clausius1879}. This fundamental inequality, besides incorporating the Carnot bound as a corollary, lays down a versatile framework for generating a plethora of more specific inequalities, such as Le Ch\^{a}telier's principle~\cite{[{For a formulation of the principle as a thermodynamic inequality, see the discussion in pages 210--214 of }] Callen1985}
and Landauer's principle~\cite{Landauer1961}. In this context, clarifying the logical connections among various universal inequalities has always been a pivotal part of thermodynamics.

In traditional thermodynamics, the scope of inequalities has been limited to those whose saturation necessitates quasistatic processes. However, in recent years, researchers have unveiled a range of thermodynamic tradeoff relations that can be saturated even during general nonequilibrium processes. These newly discovered inequalities offer nontrivial, tighter bounds on entropy production (EP) that cannot be derived from the conventional thermodynamic second law. These novel inequalities can be categorized into four distinct classes.

First, the thermodynamic uncertainty relation (TUR) states that EP is bounded from below by the ratio of a function of some mean current-like observable to the variance of the same observable. Initially discovered for biomolecular processes~\cite{Barato2015}, soon the TUR has been derived for Markov jump systems~\cite{Gingrich2016, Horowitz2017, Liu2020, Shiraishi2021}, both overdamped~\cite{Dechant2018A, Hasegawa2019, Koyuk2020, park2021} and underdamped Langevin systems~\cite{Lee2021, Kwon2022}, and even open quantum systems~\cite{Hasegawa2020, Hasegawa2021, VanVu2022}. In the steady state without any reversible currents, the lower bound of the TUR has a particularly simple form (the inverse of the relative fluctuation); but, the original form of the TUR fails in other cases, such as systems in the underdamped regime~\cite{Pietzonka2022} or with time-dependent protocols, and the inequality typically involves nontrivial derivatives of the mean current whose empirical measurement is not always straightforward.

Second, the entropic bound (EB) establishes a lower bound for EP based on the square of the mean value of a current-like observable. While this bound has been demonstrated for both overdamped and underdamped Langevin systems~\cite{Dechant2018B}, its validity has not been proven for Markov jump systems. The EB differs from the TUR in that the EB's lower bound is solely expressed in terms of the mean current without any derivatives applied to it for any nonsteady processes. This characteristic may make the EB more accessible to empirical verification, although it comes at the expense of losing an explicit upper bound on the precision of the current-like observable. As a result, the TUR and the EB have been treated as independent inequalities, without either being shown to be more fundamental than the other.

Third, the power--efficiency (PE) tradeoff shows that the power of a heat engine cannot exceed a quadratic function of its efficiency. This upper bound reaches zero at both the Carnot and zero efficiency points, precluding the existence of a finite-size heat engine that sustains positive power at Carnot efficiency. Initially, the PE tradeoff was proven as an independent inequality, unrelated to the TUR or the EB~\cite{Brandner2015, Shiraishi2016}.

Lastly, the classical speed limit (CSL) states that the EP rate is bounded from below by the square of the speed at which the system crosses the ``distance'' between the initial and the final probability distributions~\cite{Shiraishi2018, VanVu2021}. Initially conceived as the classical counterpart of the quantum speed limit (QSL)~\cite{Deffner2017}, recent studies formulate the CSL using the measure of distance between probability distributions borrowed from information geometry and optimal transport theory~\cite{Nakazato2021, Dechant2022, VanVu2023}. Thus, the CSL provides a novel geometric perspective on the issues of irreversibility and dissipation.


The crucial inquiry pertains to whether these tradeoff relations are distinct, independent inequalities, or if they can be incorporated within a single unified theoretical framework. Indeed, numerous studies have tried to establish connections among these relations. For instance, in the case of a steady-state heat engine, the PE tradeoff can be derived using the TUR~\cite{Pietzonka2018}. Similarly, for a heat engine described by Langevin equations with a time-dependent protocol, the PE tradeoff can be viewed as a consequence of the EB~\cite{Dechant2018B}.
Efforts have also been made to find a unified framework for the TUR and the CSL. A recent study derived the CSL from a short-time version of the unified thermodynamic-kinetic uncertainty relation~\cite{Vo2022}, yet the derivation's applicability is limited to step-wise or time-independent protocols, thus making the derivation not applicable to general time-dependent protocols, which are of practical interest. Other works, such as Refs.~\cite{Vo2020, Hasegawa2023}, proposed a general information-geometric inequality that encompasses the TUR and the CSL as its consequence. However, the resulting TUR in Ref.~\cite{Vo2020} is a variant version called the exponential TUR, which provides much looser bound compared to the conventional TUR. Meanwhile, Ref.~\cite{Hasegawa2023} reveals a kinetic uncertainty relation (KUR) rather than the conventional TUR. Although Ref.~\cite{Lee2023} verified the equivalence between the short-time TUR and the EB, there has been no discussion concerning the finite-time TUR and the EB. To summarize, all previous attempts to unify the tradeoff relations have been carried out under certain restrictions. A theory that successfully unifies all these relations across general settings remains elusive.



\begin{figure*}
	\includegraphics[width=0.99\textwidth]{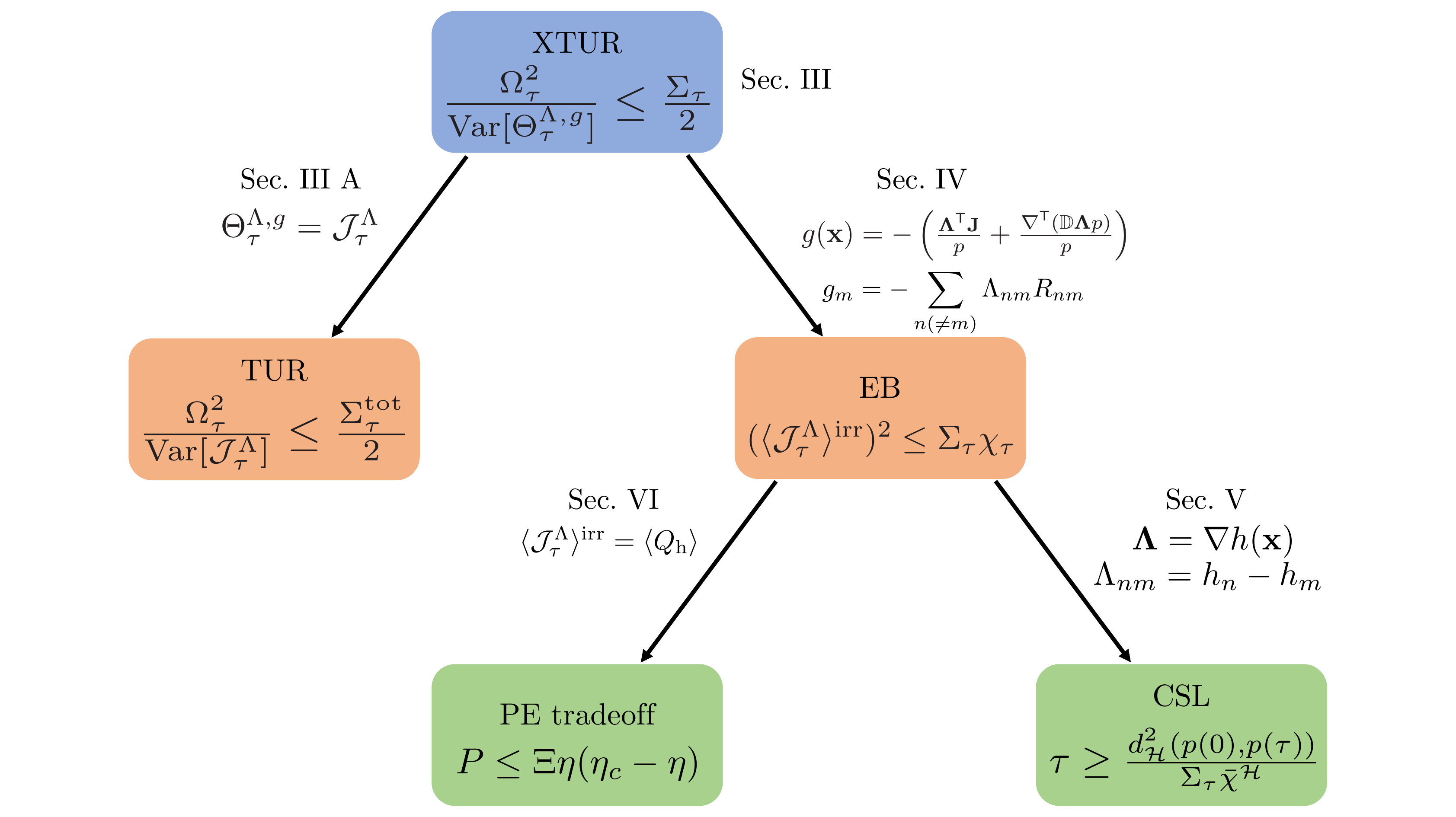}
    \caption{\label{fig:fig1} The hierarchical relationship among various thermodynamic tradeoff relations. Each box corresponds to an inequality, and each arrow illustrates the hierarchical connection between the inequalities it joins, along with the associated information specifying the conditions for deducing lower-level bounds. 
    }
\end{figure*}

In this study, we establish a unified hierarchical framework encompassing all tradeoff relations for both Langevin and Markov jump systems. This is achieved by deriving an extended version of the TUR (XTUR) applicable to both types of systems. Then, as depicted in Fig.~\ref{fig:fig1}, all the inequalities listed above are connected by a hierarchical structure originating from the XTUR. In this hierarchy, the higher-level inequalities are applicable to a broader range of observables, whereas more specific assumptions about the observables are required to derive the lower-level inequalities. Of particular significance is the central role of the EB in this hierarchical framework, serving as the intermediary between the XTUR and the lowest-level tradeoff relations. Furthermore, this hierarchical structure provides a useful guide for establishing novel thermodynamic inequalities, such as the EB for Markov jump systems and a generalized form of the CSL.

The rest of this paper is organized as follows. First, we introduce the notations used in this paper and summarize the main results (Sec.~\ref{sec:main}). Then, we prove the XTUR for the most general types of observables (Sec.~\ref{sec:XTUR}). This is followed by the derivations of the EB from the XTUR (Sec.~\ref{sec:XTUR2EB}), the generalized CSL from the EB (Sec.~\ref{sec:EB2CSL}), and the PE tradeoff from the EB (Sec.~\ref{sec:EB2PE}). After presenting concrete examples of the XTUR, the EB for Markov jump systems, and the generalized CSL (Sec.~\ref{sec:examples}), we conclude with a discussion of the possible future works (Sec.~\ref{sec:conclusions}).

\section{Notations and the main results} \label{sec:main}

In this section, we present the key thermodynamic tradeoff relations and their hierarchical relationship applicable to both Langevin and Markov jump systems. To facilitate this exploration, we begin by introducing the notations utilized throughout the paper.

\subsection{Notations for Langevin systems}

We consider an $N$-dimensional system whose state is described by the vector $\mathbf{x}=(x_1, x_2, \cdots, x_N)^{\textsf T}$. Given a drift force $\mathbf{A} = (A_1, A_2, \cdots, A_N)^{\textsf T}$ and a Gaussian white noise $\boldsymbol\xi = (\xi_1,\xi_2,\cdots,\xi_N)^{\textsf T}$, the Langevin equation can be written as 
\begin{align} \label{eq:Langevin_eq}
    \dot{\mathbf{x}}(t) = \mathbf{A}(\mathbf{x},\omega t) + \mathbb{B}(\mathbf{x}, \omega t) \bullet \boldsymbol\xi (t)
    \;,
\end{align}
where $\omega$ is a parameter controlling the speed of the protocol, $\bullet$ denotes the It\^o product, the noise satisfies $\langle \xi_i (t) \rangle=0$ and $\langle \xi_i(t) \xi_j (t') \rangle = \delta_{ij} \delta(t-t')$, and $\frac{1}{2} \mathbb{B}(\mathbf{x}, \omega t)\,\mathbb{B}^{\textsf T}\!(\mathbf{x}, \omega t) \equiv \mathbb{D}(\mathbf{x}, \omega t) $ is a position-dependent diffusion matrix. The drift force can be divided into reversible and irreversible parts as $A_i = s f_i ^\text{rev}( \mathbf{x},\omega t) + f_i ^\text{irr}( \mathbf{x},\omega t)$, where $s$ is a parameter controlling the magnitude of the reversible drift. The two force components satisfy
\begin{align} 
    f_i ^\text{rev}(\boldsymbol\epsilon \mathbf{x},\omega t) &= -\epsilon_i f_i ^\text{rev}(\mathbf{x},\omega t)\;,
    \nonumber\\ f_i ^\text{irr}(\boldsymbol\epsilon \mathbf{x},\omega t) &= \epsilon_i f_i ^\text{irr}(\mathbf{x},\omega t)
    \;,
\end{align}
where we use the notation $\boldsymbol\epsilon \mathbf{x} \equiv (\epsilon_1 x_1,\epsilon_2 x_2,\cdots,\epsilon_N x_N)$, with $\epsilon_i$ being the parity of variable $x_i$ under time reversal. 
Note that $s$ and $\omega$ are auxiliary variables that can be set to unity in the final step.

The equivalent Fokker-Planck equation is
\begin{align} \label{eq:FP}
    \partial_t p(\mathbf{x},t;s,\omega) = - \nabla^{\textsf T}  \left[\mathbf{J}^{\text{rev}}(\mathbf{x},t;s,\omega) + \mathbf{J}^{\text{irr}}(\mathbf{x},t;s,\omega)\right]
    \;,
\end{align}
where $\mathbf{J}^{\text{rev}}$ and $\mathbf{J}^{\text{irr}}$ are reversible and irreversible probability currents defined as 
\begin{align} 
    &\mathbf{J}^{\text{rev}} = s \boldsymbol{f}^{\text{rev}}(\mathbf{x},\omega t)\, p(\mathbf{x},t;s,\omega)\;,
    \nonumber \\ &\mathbf{J}^{\text{irr}} = \left\{ \boldsymbol{f}^{\text{irr}}(\mathbf{x},\omega t) - \left[\nabla^{\textsf T} \mathbb{D} (\mathbf{x},\omega t) \right ]^{\textsf T} \right\} p(\mathbf{x},t;s,\omega)
    \;,
\end{align}
respectively, with the gradient $\nabla \equiv (\partial_{x_1}, \partial_{x_2}, \cdots, \partial_{x_N})^{\textsf T}$. The total probability current is then $\mathbf{J} \equiv \mathbf{J}^{\text{rev}}+\mathbf{J}^{\text{irr}}$.
With these, the mean total EP can be written as
\begin{equation} \label{eq:total_EP_Langevin}
    \Sigma_\tau^{\rm tot} =  \int_0^\tau dt \int d\mathbf{x} \; \frac{ {\mathbf{J}^{\rm irr}}^\textsf T \mathbb{D}^{-1} (\mathbf{x},\omega t)  \mathbf{J}^{\rm irr}  }{p(\mathbf{x},t;s,\omega)}.
\end{equation}
For the brevity of notation, we shall henceforth omit the function arguments when there is no risk of confusion.

In this study, we consider both current-like and state-dependent observables, which take the forms
\begin{align}\label{eq:general_observable_Langevin}
    &\mathcal{J}_{\tau}^{\Lambda} (\Gamma) \equiv \int_0 ^\tau dt \, \boldsymbol\Lambda^{\textsf T}\!(\mathbf{x} (t),\omega t;s) \circ \dot{\mathbf{x}} (t) \;,
    \nonumber\\ &\mathcal{K}_{\tau}^g(\Gamma) \equiv \int_0 ^\tau dt \, g(\mathbf{x} (t),\omega t;s)
    \;,
\end{align}
respectively. Here $\Gamma \equiv \{ \mathbf{x}(t) \}$ denotes a stochastic trajectory, and $\circ$ stands for the Stratonovich product. In general, one can measure the composite observable $\Theta_\tau^{\Lambda,g} (\Gamma)$ involving these two: 
\begin{equation} \label{eq:composite_obs}
\Theta_\tau^{\Lambda,g} (\Gamma) \equiv \mathcal{J}_\tau^{\Lambda}(\Gamma) + \mathcal{K}_\tau^{g}(\Gamma) \;. 
\end{equation}
Using the identity $\langle \boldsymbol\Lambda^{\textsf T} \circ \dot{\mathbf{x}} \rangle = \int d\mathbf{x} \, \boldsymbol\Lambda^{\textsf T} {\textbf J} $, the mean current-like observable can be separated into reversible and irreversible components as
\begin{equation}
    \langle \mathcal J_\tau^\Lambda  \rangle = \int_0^\tau dt \int d\mathbf{x} \,\boldsymbol\Lambda^{\textsf T} \mathbf{J} = \langle \mathcal J_\tau^\Lambda  \rangle^{\rm rev}+ \langle \mathcal J_\tau^\Lambda  \rangle^{\rm irr} \;,
\end{equation}
where $\langle \mathcal J_\tau^\Lambda  \rangle^{\rm rev} \equiv \int_0^\tau dt \int d\mathbf{x} \,\boldsymbol\Lambda^{\textsf T} \mathbf{J}^{\rm rev}$ and $\langle \mathcal J_\tau^\Lambda  \rangle^{\rm irr} \equiv \int_0^\tau dt \int d\mathbf{x} \,\boldsymbol\Lambda^{\textsf T} \mathbf{J}^{\rm irr}$.

\subsection{Notations for Markov jump systems}

Here we consider a continuous-time Markov jump process within a discrete state space $S$. The transition rate from state $m$ to $n$ at time $t$ is denoted by $R_{nm} (\omega t)$, where $\omega$ is the speed of the protocol. Then, the master equation of the system is given by 
\begin{align} \label{eq:Markov_ori_dyna}
    \dot p_n(t;\omega) = \sum_{m (\neq n)} j_{nm}(t; \omega) 
    \;,
\end{align}
where $p_n(t;\omega)$ is the probability of finding the system in state $n \in S$ at time $t$, and $j_{nm}(t; \omega) =  R_{nm} (\omega t) p_m(t;\omega) -  R_{mn} (\omega t) p_n(t;\omega)$ is the net probability current from state $m$ to $n$. The mean total EP of this system is given by
\begin{equation} \label{eq:total_EP_Markov}
    \Sigma_\tau^{\rm tot} = \int_0^\tau dt \sum_{n > m} j_{nm}  \ln \frac{R_{nm}p_m  }{R_{mn}p_n }\;.
\end{equation}
Moreover, we can define the mean pseudo-EP
\begin{align} \label{eq:ps_EP_Markov}
    \Sigma_\tau ^{\text{ps}}  \equiv \int_0 ^\tau dt \sum_{n \neq m} \frac{j_{nm} ^2}{a_{nm}}
    \;,
\end{align}
where $a_{nm} = R_{nm}p_m  + R_{mn}p_n $ is the dynamical activity between states $m$ and $n$. Using the log-mean inequality, one can prove 
\begin{equation} \label{eq:ps_tot}
    \Sigma^{\text{ps}}_\tau \le \Sigma^{\text{tot}}_\tau \,.
\end{equation}
It is worth noting that, by taking the suitable continuum limit of a Markov jump process on the lattice, $\Sigma_\tau^{\rm ps}$ can also be defined for Langevin dynamics; however, the quantity coincides with $\Sigma_\tau^{\rm tot}$~\cite{Shiraishi2021}, making their distinction meaningful only for Markov jump processes.

Consider a trajectory of this Markov jump system, through which a total of $M$ jumps occur during the time interval $[0, \tau]$, with the $i$-th jump taking place at time $t_i$ for $i=1,2,\cdots,M$.
For this trajectory, let us define $N_{nm}(t)$, the number of jumps from state $m$ to $n$ up to time $t$, and $\eta_n (t) \equiv \delta_{x(t),n}$, where $x(t)$ is the state of the system at time $t$. Note that these two observables satisfy $\langle \dot{N}_{nm} (t) \rangle = R_{nm} (\omega t) p_m (t;\omega)$ and $\langle \eta_n \rangle= p_n (t;\omega)$. Then, the current-like and the state-dependent observables can be written as
\begin{align}\label{eq:general_observable_Markov}
    &\mathcal{J}_{\tau}^\Lambda (\Gamma) \equiv \int_0 ^\tau dt \sum_{n \neq m}  \Lambda_{nm} (\omega t)  \dot{N}_{nm} (t)\;,~~~(\Lambda_{nm}=-\Lambda_{mn})
    \nonumber\\ &\mathcal{K}_{\tau}^g(\Gamma) \equiv \int_0 ^\tau dt \sum_n \eta_n (t) g_n (\omega t)
    \;,
\end{align}
respectively. Just as in the case of Langevin systems, one can measure the composite observable $\Theta_\tau^{\Lambda,g}(\Gamma) \equiv \mathcal{J}_\tau^\Lambda(\Gamma) + \mathcal{K}_\tau^g (\Gamma)$.

\subsection{Main results}

Our primary finding is a unified hierarchical framework integrating the four major thermodynamic tradeoff relations (TUR, EB, CSL, and PE tradeoff) applicable to both Langevin and Markov jump systems. This hierarchy is depicted in Fig.~\ref{fig:fig1}. Each arrow in the figure, accompanied by its specifics, indicates how the lower-level inequality is deduced from the the one above it by constraining the associated observable. The reader is cautioned that, throughout this study, all expressions are given in units where the Boltzmann constant is unity.

At the apex of the hierarchy is the extended thermodynamic uncertainty relation (XTUR), which serves as the thermodynamic tradeoff relation for the most general class of the composite observable $\Theta_\tau^{\Lambda,g}(\Gamma)$. It is formulated as
\begin{align} \label{eq:XTUR1}
    \frac{\Omega_\tau ^2}{\text{Var}[\Theta_\tau^{\Lambda,g}]} \le \frac{ \Sigma_\tau }{2}
    \;,
\end{align}
where $\Sigma_\tau$ can be either $\Sigma_\tau^{\rm tot}$ or $\Sigma_\tau^{\rm ps}$ (as previously noted, they are the same quantity for Langevin systems), and $\Omega_\tau$ is defined as
\begin{align} \label{eq:omega_def}
    \Omega_\tau \equiv \hat O_\tau \langle \Theta_\tau^{\Lambda,g} \rangle - \langle \mathcal{K}_\tau^g \rangle + 
    \left\langle \Theta_\tau^{s\partial_s \Lambda,s\partial_s g} \right\rangle\;, 
\end{align}
with the operator $\hat O_t \equiv t\partial_t - s\partial_s - \omega \partial_\omega$ for both Langevin and Markov jump systems. We note that the terms associated with $s \partial_s$ in Eq.~\eqref{eq:omega_def} vanish for systems with only even-parity variables since $s=0$ there. A derivation of Eq.~\eqref{eq:XTUR1} is given in Sec.~\ref{sec:XTUR}. 


We stress that the XTUR represents the most comprehensive version of the TUR in the following sense: it provides an upper bound on the precision of the general composite observable even when the system involves odd-parity variables (including both momentum-like and force-like~\cite{Shankar2018} ones) and time-dependent protocols.
In contrast, prior research has predominantly focused on TURs for only current-like observables. While a handful of studies have ventured into deriving TURs for composite observables~\cite{Dechant2021A,Shiraishi2021,Dechant2021B}, their outcomes are restricted to the steady state and even-parity variables.



Starting from the XTUR, two other tradeoff relations are derived by restricting the form of the state-dependent observable. First, the conventional TUR for only current-like observables is readily obtained by setting the state-dependent observable to zero ($g = 0$) within the XTUR, {\em i.e.}, $ \Theta_\tau^{\Lambda,g} (\Gamma) = \mathcal J_\tau^\Lambda (\Gamma)$ as shown in Fig.~\ref{fig:fig1} and Sec.~\ref{sec:XTUR}.
Alternatively, by selecting a state-dependent observable satisfying 
\begin{align} 
    g(\mathbf{x}) &= - \left[\frac{\boldsymbol\Lambda^{\textsf T}  \mathbf{J}}{p} + \frac{\nabla^{\textsf T}  (\mathbb{D}\boldsymbol\Lambda p)}{p} \right] & \textrm{(Langevin),} \label{eq:q_EB}   \\
    g_m &= - \sum_{n \neq m} \Lambda_{nm} R_{nm} & \textrm{(Markov jump),}
\end{align}
where $p=p(\mathbf{x},t;s,\omega)$ is the probability density function, then the XTUR reduces to the EB
\begin{align}\label{eq:EB}
    (\langle \mathcal{J}_\tau^\Lambda \rangle^{\rm irr} )^2 \le \Sigma_\tau \chi_\tau
    \;,
\end{align}
where $\chi_\tau = \int_0 ^\tau dt \int d\mathbf{x} \langle \boldsymbol\Lambda ^{\textsf T} \mathbb{D} \boldsymbol\Lambda \rangle $ for Langevin systems and $\chi_\tau =\frac{1}{2} \int_0 ^\tau dt \sum_{n< m} \Lambda_{nm} ^2 a_{nm}$ for Markov jump systems.
This clarifies the relationship between the TUR and the EB: they are two different manifestations of a common higher-level tradeoff relation, the XTUR. 

It is worth noting that, compared to the TUR, the EB offers a more experimentally accessible approach for estimating EP~\cite{Lee2023}. This can be attributed to the following two reasons. First, the EB inequality can always be saturated for arbitrary nonequilibrium processes, whereas this is not generally true for the TUR, as we discuss in Sec.~\ref{sec:XTUR}. Thus, while the accurate EP estimation can be made in the EB framework, the same level of accuracy is not guaranteed for the TUR method. Second, the EB lacks any differential operators, which the TUR must incorporate in the presence of odd-parity variables or time-dependent protocols. Empirical measurements of derivatives are often challenging, if not practically unfeasible. Thus, when dealing with time-dependent protocols or odd-parity variables, the EB offers a more feasible experimental framework than the TUR for EP estimation.


Now, proceeding further down the hierarchy, the absence of differential operators in the EB facilitates connections with the two lower-level thermodynamic tradeoff relations. The first of these is the CSL. By focusing on the overdamped Langevin systems with $\boldsymbol\Lambda = \nabla h(\mathbf{x})$ or the Markov jump systems with $\Lambda_{nm} = h_n - h_m$, and taking the supremum of Eq.~\eqref{eq:EB} over a function space $ \mathcal{H}$ of $h$, we obtain the following CSL:
\begin{align}\label{eq:speedlimit}
    \tau \ge \frac{d _{\mathcal{H}}^2(p(0), p(\tau))}{ \Sigma_\tau \bar{\chi} ^{\mathcal{H}}}
    \;,
\end{align}
where $d_{\mathcal{H}}(p, q)$ is the integral probability metric (IPM), which is a measure of distance between two distributions $p$ and $q$, and $\bar{\chi} ^{\mathcal{H}}$ is the time average of the supremum of $\chi_\tau$. See Sec.~\ref{sec:EB2CSL} for details.
From Eq.~\eqref{eq:speedlimit}, we can derive a broad range of CSLs reported in the literature, such as the original CSL for Markov jump systems, Eq.~\eqref{eq:CSL_Shiraish}~\cite{Shiraishi2018}, the CSL involving the Wasserstein distance,  Eq.~\eqref{eq:CSL_Wasserstein}, and the tighter CSL, Eq.~\eqref{eq:speed_limit_lee}~\cite{Lee2022}. 


Finally, the other lower-level corollary of the EB is the PE tradeoff relation. For the case of Langevin systems, Dechant and Sasa have demonstrated that the PE tradeoff can be deduced from the EB for both steady-state and cyclic heat engines~\cite{Dechant2018B}. Meanwhile, Pietzonka and Seifert derived the PE tradeoff from the TUR, but solely for steady-state heat engines~\cite{Pietzonka2018}. As for cyclic heat engines, while some links between the TUR and the PE tradeoff have been discussed~\cite{Koyuk2019}, a close analog of the original PE tradeoff has not been derived from the TUR. This is primarily due to the inclusion of differential operators in the TUR formulation. In Sec.~\ref{sec:EB2PE}, we show that the EB makes it possible to derive the PE tradeoff for Markov jump systems, encompassing both steady-state and cyclic heat engines.


\section{XTUR} \label{sec:XTUR}

In this section, we present a derivation of the XTUR, situated at the apex of the hierarchy, for both Langevin and Markov jump systems. Following that, we delve into the equality condition of the XTUR, showing that it is achievable for the general nonequilibrium processes. Finally, we discuss the connection between the XTUR and the TUR.

\subsection{XTUR for Langevin systems} \label{sec:XTUR_A}

Adapting a previously developed method~\cite{Hasegawa2019,Lee2021}, the XTUR is derived by applying the Cram\'{e}r--Rao inequality in conjunction with the perturbed dynamics described by the following Fokker-Planck equation:
\begin{align} \label{eq:modified_Langevin}
    \partial_t p_\theta  = - \nabla^{\textsf T}  \left[\mathbf{J}_\theta^{\text{rev}} + (1+\theta)\mathbf{J}_\theta^{\text{irr}}\right]
    \;,
\end{align}
where $\mathbf{J}_\theta^{\text{rev}}$ and $\mathbf{J}_\theta^{\text{irr}}$ are given by 
\begin{align}
    \mathbf{J}_\theta^{\text{rev}} = s \mathbf{f}^{\text{rev}}p_\theta\;,\, \quad\mathbf{J}_\theta^{\text{irr}} = \left( \mathbf{f}^{\text{irr}} - \left[\nabla^{\textsf T} \mathbb{D} \right ]^{\textsf T} \right) p_\theta \,.
\end{align}
In other words, the evolution of $p_\theta$ is achieved by multiplying the perturbation factor $1+\theta$ solely to the irreversible component of Eq.~\eqref{eq:FP}. It is straightforward to check that the solution of Eq.~\eqref{eq:modified_Langevin} is related to that of Eq.~\eqref{eq:FP} by $p_\theta (\mathbf{x},t;s,\omega) = p(\mathbf{x}, t_\theta; s_\theta, \omega_\theta)$, where $t_\theta \equiv (1+\theta)t$, $s_\theta \equiv s/(1+\theta)$, and $\omega_\theta \equiv \omega / (1+\theta)$. Note that this scaling perturbation amounts to changing the drift vector $\mathbf{A}$ to $\mathbf{A}_\theta$ in the original Langevin dynamics described by Eq.~\eqref{eq:Langevin_eq}  as follows:
\begin{align} \label{eq:modified_drift}
    \dot{\mathbf{x}} = \mathbf{A}_\theta + \mathbb{B} \bullet \boldsymbol\xi \;,~
    ~~{\rm where }~~\mathbf{A}_\theta  \equiv \mathbf{A} + \theta \frac{\mathbf{J}_\theta^{\text{irr}}}{p_\theta}
    \;.
\end{align}

Within this perturbation scheme, the XTUR is derived by taking $\theta = 0$ in the Cram\'{e}r--Rao inequality
\begin{align} \label{eq:CR} 
    \frac{(\partial_\theta \langle \Theta_\tau^{\Lambda,g} \rangle_\theta)^2}{\text{Var} _\theta [\Theta_\tau^{\Lambda,g}]} \le \mathcal{I}(\theta)
    \;,
\end{align}
where $\langle \cdots \rangle_\theta$ denotes the average over trajectories of the perturbed dynamics, and $\mathcal{I}(\theta) \equiv \langle - \partial_\theta^2 \ln \mathcal{P}_\theta (\Gamma) \rangle_\theta$ represents the Fisher information associated with the path probability $\mathcal{P}_\theta (\Gamma)$ of the perturbed dynamics.

Evaluating the Fisher information at $\theta = 0$ leads to 
\begin{align} \label{eq:FI_at_0}
    \mathcal{I}(0) = \frac{1}{2} \left( \Sigma^{\text{tot}}_\tau +\mathcal B \right)
    \;,
\end{align}
where the term $\mathcal{B}$ is determined by the dependence of the initial distribution on $s$ and $\omega$ (for details, see  Appendix~\ref{secA:eval_Fisher}). 
Opting for an initial distribution dependent on $s$ or $\omega$ yields a positive $\mathcal{B}$~\cite{Lee2021}. In contrast, choosing  an initial distribution independent of $s$ and $\omega$ leads to $\mathcal{B} = 0$. There are some cases where a positive $\mathcal{B}$ leads to a tighter inequality (for a detailed explanation, see Appendix~\ref{secA:initial_dist}); however, the case of $\mathcal{B} = 0$ is more accessible to empirical measurement. Moreover, as discussed in Sec.~\ref{sec:XTUR2EB}, $\mathcal{B}=0$ allows us to derive the EB from the XTUR. Henceforth in this study, we adopt an initial distribution independent of $s$ and $\omega$, and thus, $\mathcal{B} = 0$.

Evaluation of $\partial_\theta \langle \Theta_\tau^{\Lambda,g}  \rangle _\theta \big|_{\theta=0}$ requires the calculations of both $\partial_\theta \langle \mathcal J_\tau^\Lambda  \rangle _\theta \big|_{\theta=0}$ and $\partial_\theta \langle \mathcal K_\tau^g  \rangle _\theta \big|_{\theta=0}$, which are given by
\begin{align} \label{eq:JK_calc}
    \partial_\theta \langle \mathcal J_\tau^\Lambda  \rangle _\theta \big|_{\theta=0} &= \hat O_\tau \langle \mathcal J_\tau^\Lambda \rangle +\int_0^\tau dt~ \langle (s\partial_s \boldsymbol{\Lambda}^{\textsf T} ) \circ \dot {\textbf x} \rangle   \;, \nonumber \\
    \partial_\theta \langle \mathcal K_\tau^g  \rangle _\theta \big|_{\theta=0} &= \hat O_\tau \langle \mathcal K_\tau^g \rangle - \langle \mathcal K_\tau^g \rangle + \int_0^\tau dt~ \langle s\partial_s g \rangle \;.
\end{align}
Detailed derivations are presented in Appendix~\ref{secA:evalu_J_K}. Adding these two equations results in
\begin{align} \label{eq:partial_theta}
    \partial_\theta \langle \Theta_\tau^{\Lambda,g}  \rangle _\theta \big|_{\theta=0} = \hat O_\tau \langle \Theta_\tau^{\Lambda,g} \rangle - \langle \mathcal{K}_\tau^g \rangle + 
     \left\langle \Theta_\tau^{s\partial_s \Lambda,s\partial_s g} \right\rangle , 
\end{align}
where $\Theta_\tau^{s\partial_s \Lambda,s\partial_s g} = \int_0^\tau dt \, [ (s\partial_s \boldsymbol{\Lambda}^{\textsf T} ) \circ \dot {\textbf x} + s \partial_s g ] $. 
The right-hand side of Eq.~\eqref{eq:partial_theta} is none other than $\Omega_\tau$ written in Eq.~\eqref{eq:omega_def}. 
Substituting Eq.~\eqref{eq:FI_at_0} with $\mathcal{B}=0$ and Eq.~\eqref{eq:partial_theta} into Eq.~\eqref{eq:CR}, we finally obtain the XTUR~\eqref{eq:XTUR1}.


We now turn to the issue of when the XTUR inequality saturates. Given that the XTUR is derived from the Cram\'{e}r--Rao inequality, its equality condition corresponds directly to that of the Cram\'er--Rao inequality. As shown in Eq.~(18) of Ref.~\cite{Hasegawa2019}, the equality condition is 
\begin{align} \label{eq:eq_condition}
     \Theta_\tau^{\Lambda,g} (\Gamma) - \langle \Theta _\tau^{\Lambda,g} \rangle_\theta = k(\theta) \partial_\theta \ln \mathcal{P}_\theta (\Gamma)  \;,
\end{align}
where $k(\theta)$ is an arbitrary factor. For $\theta = 0$, as detailed in  Appendix~\ref{Appendix_Equa_cond_Langevin}, this condition reduces to
\begin{align} \label{eq:equa_cond_Langevin}
    \int_0^\tau dt
    \left\{
    {\mathbf C}_1^{\textsf T}  \circ \dot{\mathbf{x}}(t)
    + {C}_2 \right\} 
    - \int_0^\tau dt \int d\mathbf{x}\, {C}_3 = 0
\end{align}
where ${\mathbf C}_1$, ${C}_2$, and ${C}_3$ are defined as
\begin{align}
    {\mathbf C}_1 
    &\equiv  \boldsymbol{\Lambda}
    - \frac{k(0)}{2} \frac{\mathbb{D}^{-1} \mathbf{J}^\text{irr}}{p} ~, \nonumber\\
    {C}_2  
    &\equiv  g + \frac{k(0)}{2} \left(\frac{{\mathbf{J}^\text{irr}}^\mathsf{T}\mathbb{D}^{-1}\mathbf{J}}{p^2} + \frac{\nabla^{\textsf T} \mathbf{J}^\text{irr}}{p}
    \right ), \nonumber \\
    {C}_3 
    &\equiv  \boldsymbol{\Lambda}^{\textsf T} \mathbf{J} + g p ~.
\end{align}
Since Eq.~\eqref{eq:equa_cond_Langevin} must hold for all trajectories, the equality condition is satisfied only when $\textbf{C}_1= {C}_2 ={C}_3 = 0$. The condition for $\mathbf{\Lambda}$ and $g$ to achieve $\textbf{C}_1 = 0$ and ${C}_2 = 0$ are
\begin{align}\label{eq:eq_con_LHS}
    \boldsymbol\Lambda^{\rm eq} &= \frac{k(0)}{2} \frac{\mathbb{D}^{-1}  {\mathbf{J}}^{\text{irr}} }{p } ~,
    \nonumber\\ g^{\rm eq} &= - \frac{k(0)}{2} \left[\frac{{\mathbf{J}^\text{irr}}^{\textsf T}  \mathbb{D}^{-1} {\mathbf{J}}}{p^2 } + \frac{\nabla^{\textsf T} {\mathbf{J}}^{\text{irr}}}{ p } \right]
    \;.
\end{align}
Interestingly, these $\boldsymbol\Lambda^{\rm eq}$ and $g^{\rm eq}$ also make $\int d{\textbf x}\,C_3$ vanish.
Thus, Eq.~\eqref{eq:eq_con_LHS} is the equality condition of the XTUR. It is worth noting that the equality condition cannot be met in the conventional TUR only with current-like observables ({\em i.e.}, $g=0$), as $g^{\rm eq}$ is generally nonzero.

Finally, we discuss the connection between the XTUR and the previously discovered TURs. First, for overdamped systems without odd-parity variables, we can set $s = 0$. For these systems, if we consider only current-like observables, then $g = 0$. In this case, the XTUR is reduced to the Koyuk--Seifert TUR~\cite{Koyuk2020}
\begin{align} 
    \frac{[(\tau \partial_\tau - \omega \partial_\omega)\langle \mathcal{J}_\tau^\Lambda \rangle]^2}{\text{Var}[\mathcal{J}_\tau^\Lambda ]} \le \frac{\Sigma^{\text{tot}}_\tau}{2}
    \;,
\end{align}
which holds for overdamped processes driven by any time-dependent protocol. Alternatively, if we set $g= - \boldsymbol{\Lambda}^{\textsf T} {\mathbf J}/p$ for overdamped processes, the XTUR in the steady state becomes the relative TUR (RTUR) proposed by Dechant and Sasa~\cite{Dechant2021A}, which is given by
\begin{align} 
    \frac{\langle \mathcal{J}_\tau^\Lambda \rangle_{\text{ss}} ^2}{\text{Var} _{\text{ss}}(\mathcal{J}_\tau^\Lambda + \mathcal{K}_\tau^g)}\le \frac{\Sigma^{\text{tot}}_\tau}{2}
    \;.
\end{align}

The XTUR can also be simplified to the TURs for underdamped Langevin dynamics. For the $N$-dimensional underdamped dynamics with position $\mathbf{x} = (x_1,\cdots,x_N )^{\textsf T}$ and momentum $\mathbf{p} = (p_1,\cdots,p_N )^{\textsf T}$,  Eq.~\eqref{eq:Langevin_eq} can be rewritten as
\begin{align} \label{eq:underdamped_Langevin_eq}
    &\dot{x}_i = s \frac{p_i}{m_i} \;, \nonumber \\
    & \dot{p_i} = s  f_i ^{\text{rev}} (\mathbf{x},\mathbf{p},\omega t) + f_i ^{\text{irr}} (\mathbf{x},\mathbf{p},\omega t) - \frac{\gamma_i}{m_i} p_i + \xi_i  \;,
\end{align}
where $m_i$, $\gamma_i$, and $\xi_i$ are the mass, the friction coefficient, and the Gaussian white noise corresponding to component $i$, respectively, with $\langle \xi_i (t) \xi_j (t') \rangle = 2 \gamma_i T_i \delta_{ij} \delta(t-t')$. 
We note that, in the underdamped dynamics, the diffusion matrix $\mathbb{D}$ is singular; however, one can directly apply our approach by simply replacing $\mathbb{D}^{-1}$ with the Moore--Penrose pseudoinverse.
For the current-like observable defined as
\begin{align} 
    \mathcal{J}_\tau^{\Lambda} (\Gamma) = \int_0 ^\tau dt\, \boldsymbol\Lambda (\mathbf{x},\mathbf{p},\omega t) \circ \dot{\mathbf{x}}
    \;,
\end{align}
it is straightforward to show that the XTUR reduces to the underdamped TUR
\begin{align} \label{eq:underdampedTUR}
    \frac{\Omega_\tau^2}{\text{Var}[\mathcal{J}_\tau^\Lambda]} \le \frac{\Sigma^{\text{tot}}_\tau}{2}
    \;,
\end{align}
where $\Omega_\tau = \hat O_\tau \langle \mathcal{J}_\tau^{\Lambda} \rangle $. While this expression is analogous to the result of Ref.~\cite{Lee2021}, there are a couple of notable differences. First, the term $\mathcal B$, dependent on the initial distribution, is absent due to our choice of an initial distribution independent of $s$ and $\omega$, as discussed earlier in this section.
Second, while $\Omega_\tau$ in Ref.~\cite{Lee2021} involves a partial derivative with respect to the spatial rescaling parameter, no such term can be found in Eq.~\eqref{eq:underdampedTUR}. This stems from the rescaling by $s$ of the inertia term $p_i/m_i$ in Eq.~\eqref{eq:underdamped_Langevin_eq}, which was not considered in Ref.~\cite{Lee2021}.

\subsection{XTUR for Markov jump systems}

The method for deriving the XTUR presented above can also be adapted to Markov jump processes. For this purpose, we consider a perturbed Markov jump dynamics characterized by a modified transition rate $R_{nm}^\theta (\omega t)$, with its corresponding master equation expressed as
\begin{equation} \label{eq:perturbed_master_eq}
    \dot{p}_n^\theta (t;\omega) = \sum_{m (\neq n)} j_{nm}^\theta (t;\omega) \,,
\end{equation}
where $j_{nm}^\theta (t;\omega) = R_{nm}^\theta (\omega t) p_m^\theta (t;\omega)- R_{mn}^\theta (\omega t) p_n^\theta (t;\omega)$. Motivated by the approaches of Refs.~\cite{Shiraishi2021,Vo2022}, we set the perturbed transition rate as $R_{nm}^\theta = R_{nm} e^{\theta K_{nm}^\theta}$ for $n \neq m$ and $R^\theta _{nn} = -\sum_{m(\neq n)} R^\theta _{mn}$ with respect to the original rate $R_{nm}$, where
\begin{align} \label{eq:K_optimal}
    K_{nm}^\theta (t; \omega) \equiv \frac{R_{nm} p_m^\theta - R_{mn} p_n^\theta}{R_{nm} p_m^\theta + R_{mn} p_n^\theta}
    \;.
\end{align}
Using the antisymmetry $K_{nm}^\theta = -K_{mn}^\theta$, we can show
\begin{equation} \label{eq:K_relation}
R_{nm} p_m^\theta  K_{nm}^\theta -  R_{mn}p_n^\theta K_{mn}^\theta = R_{nm}p_m^\theta - R_{mn}p_n^\theta. 
\end{equation}
Then, for small $\theta$, Eq.~\eqref{eq:perturbed_master_eq} can be expanded as
\begin{align} \label{eq:master_expanded}
    \dot{p}_n^\theta = &(1+\theta) \sum_{m (\neq n)}
    \left[ R_{nm}(\omega t) p_m^\theta(t;\omega) - R_{mn}(\omega t) p_n^\theta(t;\omega) \right] \nonumber \\ &+ O (\theta^2)~.
\end{align}
One can readily check that the solution of this equation is given by $p_n^\theta (t;\omega) = p_n(t_\theta;\omega_\theta) + O(\theta^2)$, where $p_n$ is the probability distribution of the original dynamics.

Similar to the approach of the previous section, in order to evaluate the Cram\'{e}r--Rao inequality~\eqref{eq:CR} at $\theta = 0$, it is necessary to compute the Fisher information, $\partial_\theta \langle \mathcal{J}_\tau^\Lambda \rangle_\theta \big|_{\theta =0}$, and $\partial_\theta \langle \mathcal{K}_\tau^g \rangle_\theta \big|_{\theta =0}$. First, choosing the $\theta$-independent initial distribution, the Fisher information at $\theta = 0$ is given by
\begin{align} \label{eq:FI_at_0_markov}
    \mathcal{I}(0) = \frac{1}{2} \Sigma_\tau^\text{ps} \leq \frac{1}{2} \Sigma_\tau^\text{tot}
    \;,
\end{align}
whose detailed derivation is presented in Appendix~\ref{sec:eval_Fisher_Markov}. Note that the inequality originates from Eq.~\eqref{eq:ps_tot}. Next, the mean current-like observable in the perturbed dynamics is obtained as
\begin{align}
    \langle \mathcal{J}_\tau^\Lambda \rangle_\theta &= \int_0 ^\tau dt \sum_{n \neq m} \Lambda_{nm} (\omega t) p^{\theta} _m (t;\omega) R^{\theta} _{nm}(t;\omega)
    \nonumber\\ &= \int_0^\tau dt \sum_{n <m} \Lambda_{nm} (\omega t) j^\theta _{nm} (t;\omega)
    \;.
\end{align}
By expanding the probability current as $j_{nm}^\theta (t;\omega) = (1+\theta) j_{nm}(t_\theta,\omega_\theta)+O(\theta^2)$, where $j_{nm}$ is the probability current of the original dynamics, we have
\begin{align}
    \langle \mathcal{J}_\tau^\Lambda \rangle_\theta &= \int_0^{\tau_\theta} dt \sum_{n <m} \ \Lambda_{nm}\left( \omega_\theta t \right) j_{nm} \left(t;\omega_\theta \right) 
    + O(\theta^2) \nonumber \\
    &=  \langle \mathcal{J}_{\tau_\theta}^\Lambda \rangle |_{\omega=\omega_\theta}
    + O(\theta^2)
    \;.
\end{align}
Therefore, we finally arrive at 
\begin{align} \label{eq:mean_J_partial_Markov}
    \partial_\theta \langle \mathcal{J}_\tau^\Lambda \rangle_\theta \big|_{\theta =0} = \hat O_\tau \langle \mathcal{J}_\tau^\Lambda \rangle 
    \;.
\end{align}
Similarly, we can prove that  \begin{align} \label{eq:mean_K_partial_Markov}
    \langle \mathcal{K}_\tau^g \rangle_\theta &= \int_0 ^\tau dt \sum_{m} g_{m} (\omega t) p^{\theta}_m (t;\omega)
    \nonumber\\ &= \frac{1}{1+\theta} \int_0^{\tau_\theta} dt \sum_{m} \  g_{m} (\omega_\theta t) p_m \left(t;\omega_\theta \right) + O(\theta^2) \nonumber \\
    &= \left. \frac{1}{1+\theta}\langle \mathcal{K}_{\tau_\theta}^g \rangle \right|_{\omega=\omega_\theta}
    + O(\theta^2)
    \;,
\end{align}
which leads to
\begin{align} \label{eq:mean_K_partial_Markov_at0}
    \partial_\theta \langle \mathcal{K}_\tau^g \rangle_\theta \big|_{\theta =0} = (\hat O_\tau -1) \langle \mathcal{K}_\tau^g \rangle 
    \;.
\end{align}
Plugging Eqs.~\eqref{eq:FI_at_0_markov}, \eqref{eq:mean_J_partial_Markov}, and \eqref{eq:mean_K_partial_Markov_at0} into the Cram\'{e}r--Rao inequality~\eqref{eq:CR} yields the XTUR~\eqref{eq:XTUR1}. Note that the terms associated with the operator $s\partial_s$ in the XTUR for Markov jump processes should vanish due to the lack of odd-parity variables.

The equality condition of the XTUR with $\Sigma_\tau = \Sigma_\tau^\mathrm{ps}$ for Markov jump systems is also identical to that of the Cram\'er--Rao inequality~\eqref{eq:eq_condition}. From Eq.~\eqref{eq:general_observable_Markov}, the left-hand side of Eq.~\eqref{eq:eq_condition} at $\theta =0$ can be written as
\begin{align} \label{eq:LHS1_Markov}
    \Theta_\tau^{\Lambda,g} &(\Gamma) - \langle \Theta _\tau^{\Lambda,g} \rangle = \int_0^\tau dt \left( 
    \sum_{n \neq m} \Lambda_{nm} \dot{N}_{nm} + \sum_n \eta_n g_n
    \right) \nonumber \\
    &- \int_0^\tau dt \left( \sum_{n \neq m} \Lambda_{mn} p_n R_{mn} +\sum_n p_n g_n \right) \,.  
\end{align}
By using the expression for the path probability~\eqref{eq:path_markov}, the right-hand side of Eq.~\eqref{eq:eq_condition} at $\theta =0 $ can be evaluated as
\begin{align} 
    &\partial_\theta \ln \mathcal{P}_\theta (\Gamma) \big|_{\theta = 0} \nonumber \\
    &=  -\int_0 ^\tau dt \sum_{n\neq m} \eta_n  R_{mn}K_{mn}^{\theta=0} 
     + \int_0 ^\tau dt \sum_{n \neq m} K_{nm}^{\theta=0}   \dot{N}_{nm} 
    \;. 
\end{align}
Therefore, the equality condition can be rewritten as
\begin{align}\label{eq:eq_con_CS_markov}
    0&=\int_0^\tau dt \sum_{n \neq m} \dot{N}_{nm} \left\{ \Lambda_{nm} - k(0) K_{nm}^{\theta=0} \right\} \nonumber \\
    & + \int_0^\tau dt \sum_n \eta_n \left\{ g_n + k(0) \sum_{m(\neq n)} R_{mn} K_{mn}^{\theta=0} \right\} \nonumber \\
    &- \int_0^\tau dt \sum_n p_n \left( \sum_{m(\neq n)} \Lambda_{mn} R_{mn} + g_n \right) \;.
\end{align}
Given that Eq.~\eqref{eq:eq_con_CS_markov} holds for all trajectories, its every integrand must be identically zero. This condition is met when the observables satisfy $\Lambda_{nm} = \Lambda_{nm}^\mathrm{eq}$ and $g_n = g_n^\mathrm{eq}$, where
\begin{align} \label{eq:eq_con_markov}
    \Lambda_{nm}^{\rm eq}  \equiv k(0) K_{nm}^{\theta=0}, ~~~g_n^{\rm eq}  \equiv - k(0) \sum_{m (\neq n)} R_{mn} K_{mn}^{\theta =0}.
\end{align}
This is not attainable when all observables are current-like. Additionally, the condition ensures the saturation of the XTUR only when $\Sigma_\tau = \Sigma_\tau^{\rm ps}$. For the XTUR with $\Sigma_\tau = \Sigma_\tau^{\rm tot}$ to be saturated, the equality $\Sigma_\tau^{\rm ps}=\Sigma_\tau^{\rm tot}$ must hold, which is in general not fulfilled by $\Lambda_{nm}^{\rm eq}$ and $g_n^{\rm eq}$.


\section{EB from XTUR} \label{sec:XTUR2EB}
In this section, we show that the EB is derived from the XTUR by choosing a specific type of observables for both Langevin and Markov jump systems.  

\subsection{Langevin systems}

Converting the Stratonovich product to the It\^o product according to Eq.~\eqref{eqA:Ito_Strato}, we can decompose the current-like observable $\mathcal{J}_\tau^\Lambda$ into two components:
\begin{align} \label{eq:decom_of_obs}
    \mathcal{J}_\tau ^\Lambda (\Gamma) &= \int_0 ^\tau dt\, \boldsymbol\Lambda ^{\textsf T} \circ \dot{\mathbf{x}}
    \nonumber\\
    &= \int_0 ^\tau dt\, \boldsymbol\Lambda ^{\textsf T} \bullet \dot{\mathbf{x}} + \int_0 ^\tau dt\, \text{tr}\,[ \mathbb{D}\nabla \boldsymbol\Lambda^{\textsf T} ]
    \;.
\end{align}
If we set $g$ in the state-dependent observable as
\begin{align} \label{eq:gEB_expression}
    g^{\rm EB} (\mathbf{x},\omega t;s)= -\boldsymbol\Lambda^{\textsf T}  (s\mathbf{f}^{\text{rev}} + \mathbf{f}^{\text{irr}}) - \text{tr}\,[\mathbb{D}\nabla \boldsymbol\Lambda^{\textsf T} ]
    \;,
\end{align}
then the composite observable becomes
\begin{align} 
    \Theta_\tau ^{\Lambda,g^{\rm EB}}(\Gamma)  = \int_0 ^\tau dt\, \boldsymbol\Lambda ^{\textsf T}   
    \mathbb{B}(\mathbf{x}, \omega t) \bullet \boldsymbol\xi (t)
    \;.
\end{align}
Using the Gaussian statistics of $\boldsymbol\xi(t)$, it is straightforward to show that $\langle \Theta_\tau^{\Lambda,g^{\rm EB}} \rangle =0 $ and
\begin{align}  \label{eq:EBvar}
    \text{Var}(\Theta_\tau ^{\Lambda,g^{\rm EB}}) =\left\langle \left(\Theta_\tau ^{\Lambda,g^{\rm EB}}\right) ^2 \right\rangle = 2 \int_0 ^\tau dt \,\langle \boldsymbol\Lambda ^{\textsf T} \mathbb{D}\boldsymbol\Lambda \rangle
    \;.
\end{align}
In addition, using $\langle \Theta_\tau^{\Lambda,g^{\rm EB}} \rangle =0 $ in Eq.~\eqref{eq:composite_obs}, we have 
\begin{equation} \label{eq:K_-J}
    \langle \mathcal K_\tau^{g^{\rm EB}} \rangle = -\langle \mathcal J_\tau^\Lambda \rangle .  
\end{equation}
Meanwhile, using the relations $s\mathbf{f}^{\text{rev}} + \mathbf{f}^{\text{irr}} = \mathbf{J}/p+ (\nabla^{\textsf T} \mathbb{D} p)^{\textsf T} /p$ and 
\begin{equation} \label{eq:some_relation1}
    \nabla^{\textsf T} [\mathbb{D} \boldsymbol{\mathbf{V}} p] = \text{tr} \,[\mathbb{D} \nabla \mathbf{V}^{\textsf T}]\,p + \mathbf{V}^{\textsf T} (\nabla^{\textsf T}\mathbb{D} p)^{\textsf T}
\end{equation}
that holds for any $N$-dimensional vector $\mathbf{V}$, from Eq.~\eqref{eq:gEB_expression} we obtain
\begin{equation} \label{eq:s_partial_s_g}
    s\partial_s g^{\rm EB} = -\frac{(s\partial_s \boldsymbol{\Lambda}^{\textsf T}) \mathbf{J}}{p} - \frac{1}{p} \nabla^{\textsf T}[\mathbb{D}(s\partial_s \boldsymbol{\Lambda})p] - \frac{\boldsymbol{\Lambda}^{\textsf T} \mathbf{J}^{\rm rev} }{p} .
\end{equation}
Then, utilizing Eq.~\eqref{eq:s_partial_s_g}, the third term in Eq.~\eqref{eq:omega_def} is evaluated as
\begin{align} \label{eq:third_term}
    \left\langle \Theta_\tau^{s\partial_s \Lambda,s\partial_s g^{\rm EB}} \right\rangle &=\int_0 ^\tau dt\, \int d\mathbf{x}\, (s\partial_s \boldsymbol\Lambda^{\textsf T}  \mathbf{J} + s\partial_s g^{\rm EB}  p) \nonumber
    \\ &=- \int_0 ^\tau dt\, \int d\mathbf{x}\, \boldsymbol\Lambda^{\textsf T}  \mathbf{J}^{\text{rev}}
    \;.
\end{align}
Combining Eqs.~\eqref{eq:K_-J} and \eqref{eq:third_term}, $\Omega_\tau$ in Eq.~\eqref{eq:omega_def} becomes
\begin{align} \label{eq:EB_Omega}
    \Omega_\tau 
    &=\int_0 ^\tau dt \int d\mathbf{x} \, \boldsymbol\Lambda ^{\textsf T} \mathbf{J}  - \int_0 ^\tau dt \int d\mathbf{x} \, \boldsymbol\Lambda ^{\textsf T}  \mathbf{J}^{\text{rev}}
    \nonumber\\ 
    &= \int_0 ^\tau dt \int d\mathbf{x} \, \boldsymbol\Lambda^{\textsf T} \mathbf{J}^{\text{irr}} = \langle \mathcal{J}_\tau^\Lambda \rangle^{\text{irr}}
    \;.
\end{align}
Finally, putting Eqs.~\eqref{eq:EBvar} and \eqref{eq:EB_Omega} into the XTUR yields the EB~\eqref{eq:EB}. Note that Eq.~\eqref{eq:gEB_expression} for $g^{\rm EB}$ can be transformed into the form shown in Eq.~\eqref{eq:q_EB} using Eq.~\eqref{eq:some_relation1}.

\subsection{Markov jump systems} \label{Markov jump EB}

Noting that the EB for Markov jump systems has not previously been established, we first prove it without relying on any other thermodynamic tradeoff relations. Starting with the definition of the current-like observable for Markov jump systems, shown in Eq.~\eqref{eq:general_observable_Markov}, we derive the EB as follows:
\begin{align} \label{eq:markov_EB}
    \langle \mathcal{J}_\tau^\Lambda \rangle^2
    &= \left ( \int_0 ^\tau dt \sum_{n<m} \Lambda_{nm} j_{nm} \right )^2
    \nonumber\\ &\le \frac{1}{2} \int_0 ^\tau dt \sum_{n <m}\Lambda_{nm}^2 a_{nm} \times 2\int_0 ^\tau dt \sum_{n<m} \frac{j_{nm}^2}{a_{nm}}
    \nonumber\\ &= \chi_\tau \Sigma^{\text{ps}}_\tau \le \chi_\tau \Sigma^{\text{tot}}_\tau 
    \; , 
\end{align}
where $\chi_\tau \equiv \frac{1}{2} \int_0 ^\tau dt \sum_{n <m} \Lambda_{nm}^2 a_{nm}$. For the inequalities in the second and the third lines, the Cauchy--Schwarz inequality and Eq.~\eqref{eq:ps_tot} are used, respectively.

Now, we show that Eq.~\eqref{eq:markov_EB} can be derived from the XTUR for Markov jump systems. Towards this aim,  we set the function $g$ as
\begin{align} \label{eq:gEB_expression_Markov}
    g_m^{\rm EB} = - \sum_{n (\neq m)} \Lambda_{nm} R_{nm}
    \;.
\end{align}
Then, the composite observable $\Theta_\tau^{\Lambda, g^{\rm EB}} = \mathcal{J}_\tau^\Lambda + \mathcal{K}_\tau^{g^{\rm EB}}$ takes the form
\begin{align} \label{eq:EB_Markov_observable}
    \Theta_\tau^{\Lambda,g^{\rm EB}} = \int_0 ^\tau dt \sum_{n \neq m} \Lambda_{nm} (\omega t)\, [\dot{N}_{nm}(t) - \eta_m (t) R_{nm}(\omega t)]
    \;.
\end{align}
Using the relations $\langle \dot{N}_{nm} \rangle = R_{nm} p_m$ and $\langle \eta_n \rangle= p_n$, we obtain $\langle \Theta_\tau^{\Lambda,g^{\rm EB}} \rangle = 0$, which implies $\langle \mathcal{J}_\tau^\Lambda \rangle = - \langle \mathcal{K}_\tau^{g^{\rm EB}} \rangle$. Moreover, we can derive
\begin{align} \label{eq:Var_EB}
    \text{Var}(\Theta _\tau^{\Lambda,g^{\rm EB}}) = \int_0 ^\tau dt \sum_{n \neq m} \Lambda_{nm}^2 R_{nm} p_m =2 \chi_\tau \,,
\end{align}
as detailed in Appendix~\ref{Appendix_A4.5}.
Combining these results with the XTUR~\eqref{eq:XTUR1} for $\Sigma_\tau=\Sigma_\tau^\mathrm{ps}$, we arrive at the EB as given in Eq.~\eqref{eq:markov_EB}.


\section{CSL from EB} \label{sec:EB2CSL}

In this section, we show how the CSL can be derived from the EB. To achieve this, we first introduce the integral probability metric (IPM)~\cite{Zolotarev1984,Muller1997,Sriperumbudur2012}, which quantifies the statistical difference between two probability distributions. Then the CSL is derived by applying the IPM within the EB framework.

\subsection{Integral Probability Metric (IPM)}

The IPM is a metric quantifying the \emph{distance} between two probability distributions. For any pair of distributions $p$ and $q$, the IPM is defined as
\begin{align} \label{eq:IPM_def}
    d_{\mathcal{H}} (p,q) = \sup_{h \in \mathcal{H}} \left| \langle h \rangle_p - \langle h \rangle_q \right|
    \;,
\end{align}
where $\mathcal{H}$ represents a set of functions with some specific properties, and $\langle h \rangle_p$ and $\langle h \rangle_q$ are the expectation values of the function $h$ with respect to $p$ and $q$, respectively.

There are some notable examples of the IPM. First, when $\mathcal{H}$ is a set of functions bounded within the range $[-1,1]$, the IPM is equal to twice the {\em total variation distance}~\cite{Sriperumbudur2012}, {\em i.e.}, $d_{\mathcal{H}}(p,q) = 2d_{\text{TV}}(p,q)$, where
\begin{align} \label{eq:total_var_dist}
    d_{\text{TV}} (p,q) &\equiv \frac{1}{2} \sum_{x \in \Omega} |p(x) - q(x)| \nonumber\\
    &= \frac{1}{2} \sup_{h \in [-1,1]} | \langle h \rangle_p - \langle h \rangle_q|
    \;.
\end{align}

Second, we consider the case where $\mathcal{H}$ is the set of {\em Lipschitz-$1$ functions}, defined as  $\text{Lip}_1 \equiv \{h | d(x,y) \ge |h(x)-h(y)| ,\, \forall x,y \}$, where $d(x,y)$ represents the distance between $x$ and $y$. In this case, due to the Kantorovich--Rubinstein duality~\cite{Villani2009,Santambrogio2015}, the IPM coincides with the {\em Wasserstein-$1$ distance}
\begin{align} \label{eq:d_Lip1_Wasser}
    d_{\text{Lip}_1}(p,q) = \mathcal{W}_1 (p,q)
    \;,
\end{align}
which is widely used in the context of the optimal transport problem~\cite{Villani2009,Santambrogio2015,Arjovsky2017}. For the continuous random variables, the Wasserstein-$1$ distance is defined as
\begin{align}
\mathcal{W}_1(p,q) \equiv \inf_{\pi(\mathbf{x},\mathbf{y})\in \Phi(p,q)} \int d\mathbf{x}\, d\mathbf{y}\, d(\mathbf{x},\mathbf{y}) \,\pi(\mathbf{x},\mathbf{y})
\;,
\end{align}
where $\Phi(p,q)$ stands for the set of all joint probability distributions $\pi(\mathbf{x},\mathbf{y})$ whose marginals are given by $\int d\mathbf{y}\,\pi = p$ and $\int d\mathbf{x}\,\pi = q$. For Langevin systems, $d(\mathbf{x},\mathbf{y})$ is typically taken to be the Euclidean distance $|\mathbf{x}-\mathbf{y}|$. Meanwhile, for discrete random variables, the definition changes to
\begin{align}
\mathcal{W}_1 (p, q) \equiv \inf_{\pi_{nm} \in \Phi (p, q)} \sum_{n,m} d_{nm} \pi_{nm}
\;,
\end{align}
where $\Phi(p, q)$ is the set of all joint probability distributions $\pi_{nm}$ which satisfies $\sum_{m} \pi_{nm} = p _n$ and $\sum_{n} \pi_{nm} =q _m$. For Markov jump systems, the distance $d_{nm}$ is typically set to be the length of the shortest path between states $n$ and $m$ in the graph representation of possible transitions. See Appendix~\ref{Appendix_A5} for more details about the Wasserstein-1 distance on graphs.


\subsection{CSL for Langevin systems}

Employing the IPM introduced above, the CSL~\eqref{eq:speedlimit} can be deduced from the EB~\eqref{eq:EB} by appropriately limiting the associated observable to a specific form. Since the CSL has been studied for systems without odd-parity variables, here we focus on the overdamped Langevin systems.

We start by defining $\mathcal{D}$ as the set of functions that are continuous and almost everywhere differentiable in $\mathbb{R}^N$. Let $\mathcal{H}$ be a subset of $\mathcal{D}$. Then, for any function $h \in \mathcal{H}$, $\nabla h(\mathbf{x})$ exists everywhere except for a measure-zero subset $V_h \subset \mathbb{R}^N$. With this in mind, we choose the current-like observable satisfying 
\begin{align}
    \boldsymbol\Lambda  = \begin{cases} \nabla h(\mathbf{x}),~\ & \text{if $\mathbf{x} \in \mathbb{R}^N \backslash V_h$,} \\ \boldsymbol\Lambda_0 (\mathbf{x}),~\ & \text{if $\mathbf{x} \in V_h$,} \end{cases}
\end{align}
where $\boldsymbol\Lambda_0$ is an arbitrary vector field. Then the mean value of the current-like observable is given by
\begin{align} \label{eq:J_h_Langevin}
    \langle \mathcal{J}_\tau^{\Lambda} \rangle &= \int_0^\tau dt \int_{\mathbb{R}^N \backslash V_h} d\mathbf{x}\, (\nabla^{\textsf T} h) \,\mathbf{J} + \int_0 ^\tau dt \int_{V_h} d\mathbf{x} \,\boldsymbol\Lambda_0 ^{\textsf T} \mathbf{J}
    \nonumber\\ &= \int_0 ^ \tau dt \int_{\mathbb{R}^N} d\mathbf{x}\, h(\mathbf{x})\, \partial_t p(\mathbf{x},t)\nonumber\\
    &= \langle h \rangle_{p(\tau)} - \langle h \rangle_{p(0)}
    \;,
\end{align}
where $\langle h \rangle_{p(t)} \equiv \int d\mathbf{x} \, h(\mathbf{x}) p(\mathbf{x},t) $. Integration by parts and $\partial_t p = -\nabla \cdot \mathbf{J}$ have been used to obtain the second equality, along with the fact that $V_h$ is a measure-zero set and does not contribute to the integral. Using this result in Eq.~\eqref{eq:EB} and taking the supremum over $h\in \mathcal{H}$, the EB transforms into
\begin{align} \label{eq:EB_to_CSL}
    d^2 _{\mathcal{H}} (p(0), p(\tau)) \le \Sigma^{\text{tot}}_\tau \chi^{\mathcal{H}}_\tau
    \;,
\end{align}
where $\chi^{\mathcal{H}}_\tau \equiv \sup_{h \in \mathcal{H}} \int_0 ^\tau dt \int_{\mathbb{R}^N \backslash V_h} d\mathbf{x}\,p \nabla h ^T \mathbb{D} \nabla h$ is a constant that depends on the choice of $\mathcal{H}$. Since $\chi^{\mathcal{H}}_\tau$ is a time-extensive quantity, we can define its time average $\bar{\chi} ^\mathcal{H} \equiv \chi^{\mathcal{H}}_\tau /\tau $. 
Then, Eq.~\eqref{eq:EB_to_CSL} can be rewritten in the standard CSL form as
\begin{align} \label{eq:CSL_Langevin}
    \tau \ge \frac{d _{\mathcal{H}}^2 (p(0), p(\tau))}{\Sigma_\tau^{\rm tot} \bar{\chi}^{\mathcal{H}}}
    \;.
\end{align}

By choosing a proper $\mathcal{H}$, various CSLs can be derived from this inequality. For example, let us take $\mathcal{H} = \text{Lip}_1 \subset \mathcal{D}$. Then, as stated in Eq.~\eqref{eq:d_Lip1_Wasser}, $d_{\text{Lip}_1}$ corresponds to the Wasserstein-$1$ distance. As for $\chi_\tau ^{\text{Lip}_1}$, since every $h \in \text{Lip}_1$ satisfies $|\nabla h(\mathbf{x})| \le 1$, we obtain
\begin{align} \label{eq:CSL_Lip}
    \chi_\tau ^{\text{Lip}_1} &= \sup_{h \in \text{Lip}_1} \int_0 ^\tau dt \int_{\mathbb{R}^N \backslash V_h} d\mathbf{x} \, p\nabla h ^T \mathbb{D} \nabla h 
    \nonumber\\ &\le \int_0 ^\tau dt \int d\mathbf{x} \,\rho[\mathbb{D}(\mathbf{x},\omega t)]\,p(\mathbf{x},t)
    = \int_0 ^\tau dt\,\langle \rho[\mathbb{D}] \rangle
    \;,
\end{align}
where $\rho(\mathbb{A})$ is the largest absolute eigenvalue of the matrix $\mathbb{A}$. Then, the CSL can be rewritten as
\begin{align} \label{eq:CSL_Wasserstein}
    \tau \ge \frac{\mathcal{W}_1^2 (p(0), p(\tau))}{\Sigma_\tau^{\rm tot} \overline{\langle \rho(\mathbb{D}) \rangle}}
    \;,
\end{align}
where $\overline{\langle \rho(\mathbb{D}) \rangle} \equiv \frac{1}{\tau} \int_0 ^\tau dt \,\langle \rho(\mathbb{D}) \rangle$. If the diffusion matrix is uniform in $\mathbf{x}$ and $t$, we can replace $\overline{\langle \rho(\mathbb{D}) \rangle}$ with $\rho(\mathbb{D})$. 
Or, if the diffusion matrix is an identity matrix multiplied by time-dependent temperature $T(\mathbf{x}, t)$, then $\overline{\langle \rho (\mathbb{D}) \rangle} = \frac{1}{\tau} \int_0 ^\tau dt\, \langle T(\mathbf{x},t) \rangle $, where $\langle T(\mathbf{x},t) \rangle = \int d\mathbf{x}\, T(\mathbf{x},t)\,p(\mathbf{x},t)$.

It is worth highlighting that the CSL presented in Eq.~\eqref{eq:CSL_Wasserstein}, first established by this study, is generally applicable to systems featuring a diffusion matrix which is neither diagonal nor uniform in $\mathbf{x}$ and $t$. This is distinct from the previous CSLs for overdamped Langevin systems, which focused on systems immersed a single heat bath at fixed temperature~\cite{Dechant2019, Nakazato2021, Ito2023, VanVu2023}. Additionally, we emphasize our use of the Wasserstein-$1$ distance instead of the Wasserstein-$2$ distance employed in previous works~\cite{Dechant2019, Nakazato2021, Ito2023, VanVu2023}. The difference is rooted in the fact that the Wasserstein-$1$ distance belongs to the category of the IPM, while the Wasserstein-$2$ distance does not. Simply replacing $\mathcal{W}_1$ with $\mathcal{W}_2$ in Eq.~\eqref{eq:CSL_Wasserstein} can lead to violation of the inequality, as demonstrated in Sec.~\ref{ex_CSL_Langevin}. The reader is also referred to \cite{Shiraishi2023} for a variant of the CSL which is valid for $\mathcal{W}_1$ but not for $\mathcal{W}_2$.

\subsection{CSL for Markov jump systems}

The CSL for Markov jump systems can also be derived from the EB when the associated current-like observable meets the gradient condition $\Lambda_{nm} = h_n - h_m \equiv \nabla h_{nm}$. Then, its ensemble average satisfies
\begin{align} \label{eq:nabla_h_Markov}
    \langle \mathcal{J}_\tau^{\nabla h} \rangle &= \int_0 ^\tau dt \sum_{n > m} (h_n - h_m) j_{nm}
    = \int_0 ^\tau dt \sum_{n \neq m} h_n j_{nm}
    \nonumber\\ &= \int_0 ^\tau dt \sum_n h_n \dot{p}_n (t)
    = \langle h \rangle_{p(\tau)} - \langle h \rangle_{p(0)}
    \;,
\end{align}
where $\dot{p}_n (t) = \sum_{m (\neq n)} j_{nm}$ has been used for the third equality, and $\langle h \rangle_{p(t)} \equiv \sum_n h_n p_n\!(t)$. Using this  relation, the EB~\eqref{eq:EB} for $\Sigma_\tau = \Sigma_\tau^\mathrm{ps}$ reduces to
\begin{align} \label{eq:bound_IPM}
    d^2 _{\mathcal{H}} (p(0), p(\tau)) \le \Sigma^{\text{ps}}_\tau \chi _\tau ^{\mathcal{H}}
    \;,
\end{align}
where $\chi _\tau ^{\mathcal{H}} \equiv \frac{1}{2} \sup_{h \in \mathcal{H}} \int_0 ^\tau dt \sum_{n<m} (h_n - h_m)^2 a_{nm}$. Since $\Sigma^{\text{ps}}_\tau \le \Sigma^{\text{tot}}_\tau$, once again we obtain Eq.~\eqref{eq:CSL_Langevin}, proving that the CSL for Markov jump systems can also be formulated in the same form.

For concreteness, let us examine the case of $\mathcal{H} = \mathcal{H}_1 \equiv \{ h_n | h_n \in [-1,1]\}$. Then, as discussed above Eq.~\eqref{eq:total_var_dist}, $d_\mathcal{H}$ is related to the total variation distance. Besides, since $|h_n - h_m| \le 2$ for all $n$ and $m$, we obtain
\begin{align} \label{eq:bound_chi}
    \chi_\tau^{\mathcal{H}_1} \le 2 \int_0 ^\tau dt \sum_{n <m} a_{nm} = 2\mathcal{A}_\tau
    \;,
\end{align}
where $\mathcal{A}_\tau$ is the dynamical activity. Denoting its time average as $\bar{\mathcal{A}} \equiv \mathcal{A}_\tau / \tau $, Eq.~\eqref{eq:bound_IPM} implies
\begin{align} \label{eq:CSL_Shiraish}
    \tau \ge \frac{d_{\text{TV}}^2 (p(0), p(\tau))}{\Sigma_\tau^{\rm tot} \bar{\mathcal{A}}/2}
    \;,
\end{align}
which reproduces the original CSL derived in Ref.~\cite{Shiraishi2018}. 

Another CSL, recently discovered~\cite{Lee2022} and proven to be tighter than Eq.~\eqref{eq:CSL_Shiraish}, can also be derived from Eq.~\eqref{eq:bound_IPM}. Let $f(x)$ be the inverse function of $f^{-1}(x)=x\tanh x$. Then, Eqs.~\eqref{eq:bound_IPM} and \eqref{eq:bound_chi} lead to the following inequalities: 
\begin{align} \label{eq:speed_limit_cascade}
    \frac{d_{\mathcal{H}_1} ^2 }{2 \mathcal{A}_\tau} \le \frac{d_{\mathcal{H}_1}^2 }{\chi_\tau ^{\mathcal{H}_1}} \le \Sigma^{\text{ps}}_\tau \le \frac{(\Sigma_\tau ^{\text{tot}})^2 }{2\mathcal{A}_\tau} f \left( \frac{\Sigma_\tau ^\text{tot}}{2\mathcal{A}_\tau} \right)^{-2}
    \;.
\end{align}
The final inequality can be derived using Eq.~(10) of Ref.~\cite{Vo2022}. These yield $f(\Sigma_\tau ^{\text{tot}}/2\mathcal{A_\tau}) \le \Sigma_\tau ^{\text{tot}}/d_{\mathcal{H}_1}$, which in turn implies
\begin{align} 
    \frac{\Sigma_\tau ^{\text{tot}}}{2\mathcal{A}_\tau} \le f^{-1}\left(\frac{\Sigma_\tau ^{\text{tot}}}{d_{\mathcal{H}_1}}\right) = \frac{\Sigma_\tau ^{\text{tot}}}{d_{\mathcal{H}_1}} \tanh \left(\frac{\Sigma_\tau ^{\text{tot}}}{d_{\mathcal{H}_1}}\right)
    \;.
\end{align}
Finally, using $d_{\mathcal{H}_1} =2 d_{\text{TV}}$, we arrive at
\begin{align} \label{eq:speed_limit_lee}
    \tau \ge \frac{d_{\text{TV}}}{\bar{\mathcal{A}} \tanh \left(\frac{\Sigma_\tau ^{\text{tot}}}{2d_{\text{TV}}}\right)}
    \;,
\end{align}
reproducing the tighter CSL proven in Ref.~\cite{Lee2022}.  

Yet another application of Eq.~\eqref{eq:bound_IPM} is the CSL for Markov jump systems in terms of the Wasserstein distance. To show this, we select $\mathcal{H} = \textrm{Lip}_1$, which equates $d_\mathcal{H}$ to the Wasserstein-1 distance, as stated in Eq.~\eqref{eq:d_Lip1_Wasser}. For Markov jump systems, $\textrm{Lip}_1$ can be expressed as
\begin{align} \label{eq:Lip_1_other_expression}
    \text{Lip}_1 = \left\{ h_n |  1 \ge |h_n - h_m|  \,\, \text{for all }(n,m)\in E \right\}
    \;,
\end{align}
where $E$ denotes the set of all state pairs joined by allowed transitions. A detailed proof is provided in Appendix~\ref{Appendix_A5}. Using the above property of $\text{Lip}_1$, it is straightforward to obtain
\begin{align} 
    \chi^ {\text{Lip}_1} _\tau \le \frac{1}{2} \int_0 ^\tau dt \sum_{n<m} a_{nm} = \frac{1}{2} \mathcal{A}_\tau
    \;.
\end{align}
Combining these results with Eq.~\eqref{eq:bound_IPM}, we derive
\begin{align} 
    \tau \ge \frac{\mathcal{W}_1^2 (p(0), p(\tau) )}{\Sigma_\tau^{\rm tot} \bar{\mathcal{A}}/2}
    \;.
\end{align}
As was done for the case of the original CSL discussed above, a tighter version of the above CSL, namely 
\begin{align} \label{eq:speed_limit_vu}
    \tau \ge \frac{\mathcal{W}_1 }{\bar{\mathcal{A}} \tanh \left(\frac{\Sigma_\tau^{\rm tot}}{2 \mathcal{W}_1} \right)}
    \;,
\end{align}
can be derived using inequalities similar to Eq.~\eqref{eq:speed_limit_cascade}. This result is equivalent to Eq.~(57) of Ref.~\cite{Dechant2022} and Eq.~(129) of Ref.~\cite{VanVu2023}. As discussed in \cite{VanVu2023}, Eq.~\eqref{eq:speed_limit_vu} is generally saturable for any graphs, while Eq.~\eqref{eq:speed_limit_lee} can be saturated only for fully-connected graphs.

These derivations provide clear evidence that Eq.~\eqref{eq:bound_IPM}, essentially the EB for a specific class of current-like observables, is a foundational CSL from which follows a broad range of CSLs reported in the literature.


\section{PE tradeoff from EB} \label{sec:EB2PE}

Consider a heat engine that operates between hot and cold heat reservoirs characterized by the temperatures $T_{\rm h}$ and $T_{\rm c}$ $(T_{\rm h}>T_{\rm c})$, respectively. Let $Q_{\rm h}$ denote the heat absorbed from the hot reservoir, $Q_{\rm c}$ the heat dissipated into the cold reservoir, and $W$ the work extracted from the engine. Then, the PE tradeoff relation can be expressed as~\cite{Shiraishi2016}
\begin{equation} \label{eq:PE_tradeoff} 
    P \le \Xi \eta (\eta_{\rm C} - \eta) \,,
\end{equation}
where $P$ is the power of the engine, $\Xi$ a system-dependent constant, $\eta \equiv \langle W\rangle/ \langle Q_{\rm h}\rangle$ the efficiency of the engine, and $\eta_{\rm C} = 1-T_{\rm c}/T_{\rm h}$ the Carnot efficiency. This relation dictates that it is impossible to achieve $\eta_{\rm C}$ and finite power simultaneously, even though an irreversible process can sometimes achieve $\eta_{\rm C}$~\cite{Lee2017,Lee2019}. This tradeoff relation has been proven for both steady-state and cyclic engines in Langevin systems using the EB~\cite{Dechant2018B}. It has also been derived using the TUR~\cite{Pietzonka2018}, but solely for the steady-state engines and not for the cyclic engines. In this section, we establish the PE tradeoff for both steady-state and cyclic engines in Markov jump systems using the EB.

To achieve this goal, we consider a heat engine consisting of discrete $N$ states in contact with two heat reservoirs, each characterized by a temperature denoted as $T_k$ ($k\in \{{\rm h},{\rm c}\}$). The energy associated with each state is represented as $E_n(t)$ ($n \in \{1, 2, \ldots, N\}$). The transition rate from state $n$ to state $m$ under the influence of heat reservoir $k$ is expressed as $R_{mn}^k(\omega t)$, which satisfies the local detailed balance condition: $R_{mn}^k e^{-E_n/T_k} = R_{nm}^k e^{- E_m/T_k}$. The master equation governing this system is given by
\begin{align} \label{eq:multi_bath_master_eq}
    \dot{p}_n = \sum_{m(\neq n),k} j_{nm}^k \;,
\end{align}
where $j_{nm}^k = R_{nm}^k p_m - R_{mn}^k p_n$. We are interested in the current-like observables expressed as
\begin{equation}
    \mathcal{J}_\tau^{\Delta E,k} = \int_0^\tau dt \sum_{n \neq m} \Delta E_{nm} \dot{N}_{nm}^k \,,
\end{equation}
where $\Delta E_{nm} = E_n - E_m$, and $N_{nm}^k$ is the number of jumps from state $m$ to $n$ under the influence of reservoir $k$ until time $t$. Then, we can readily identify $\mathcal{J}_\tau^{\Delta E,{\rm h}} = Q_{\rm h}$ and $\mathcal{J}_\tau^{\Delta E,{\rm c}} = -Q_{\rm c}$. Choosing $Q_{\rm h}$ as the observable appearing in the EB, we obtain
\begin{equation} \label{eq:heat_EB}
    \langle Q_{\rm h}\rangle^2 \le \tau \bar{\chi}^{\rm h} \Sigma_\tau^{\rm tot},
\end{equation}
where $\bar{\chi}^{\rm h} = \frac{1}{2 \tau} \int_0^\tau dt \sum_{n \neq m} \Delta E_{nm}^2 a_{nm}^{\rm h}$ with $a_{nm}^{\rm h} = R_{nm}^{\rm h} p_n + R_{mn}^{\rm h} p_m$. When the engine is in the periodic state, we choose $\tau$ to be the period of a cycle. Then, since the energy of the system does not change after each cycle, we have $\Sigma_\tau^{\rm tot} = \langle Q_{\rm c} \rangle /T_{\rm c} - \langle Q_{\rm h} \rangle /T_{\rm h} = (\eta_{\rm C} - \eta)\langle Q_{\rm h} \rangle/T_{\rm c}$. Using this expression in Eq.~\eqref{eq:heat_EB} and multiplying both sides of the inequality by $\eta =\langle W \rangle/\langle Q_{\rm h}\rangle$, we finally derive
\begin{equation} \label{eq:PE_Markov}
    P \le \frac{\bar{\chi}^{\rm h}}{T_{\rm c}} \eta (\eta_{\rm C} - \eta),
\end{equation}
where $P = \langle W \rangle/\tau$. It is straightforward to adapt this proof to the same inequality for steady-state engines. We stress that Eq.~\eqref{eq:PE_Markov} is an equivalent of the PE tradeoff relation shown in Eq.~\eqref{eq:PE_tradeoff}, but extended to both cyclic and steady-state engines built using Markov jump systems.

For completeness, we note that the PE tradeoff for Langevin systems can also be derived from the EB by choosing the heat from the hot reservoir as the observable of interest, as previously shown in Ref.~\cite{Dechant2018B}. Thus, the EB provides a reliable starting point for deducing the PE tradeoff for a broad range of engines. In contrast, to our knowledge, it seems impossible to derive the PE tradeoff for cyclic engines starting from the TUR.

\section{Examples} \label{sec:examples}

In this section, we present concrete illustrations of three tradeoff relations derived in this study: the XTUR for an overdamped Brownian particle pulled by optical tweezers, the EB for Markov jumps systems, and the CSL for Langevin systems subject to an inhomogeneous temperature field.

\subsection{XTUR for an overdamped Brownian particle pulled by optical tweezers}

\begin{figure}
    \includegraphics[width=\columnwidth]{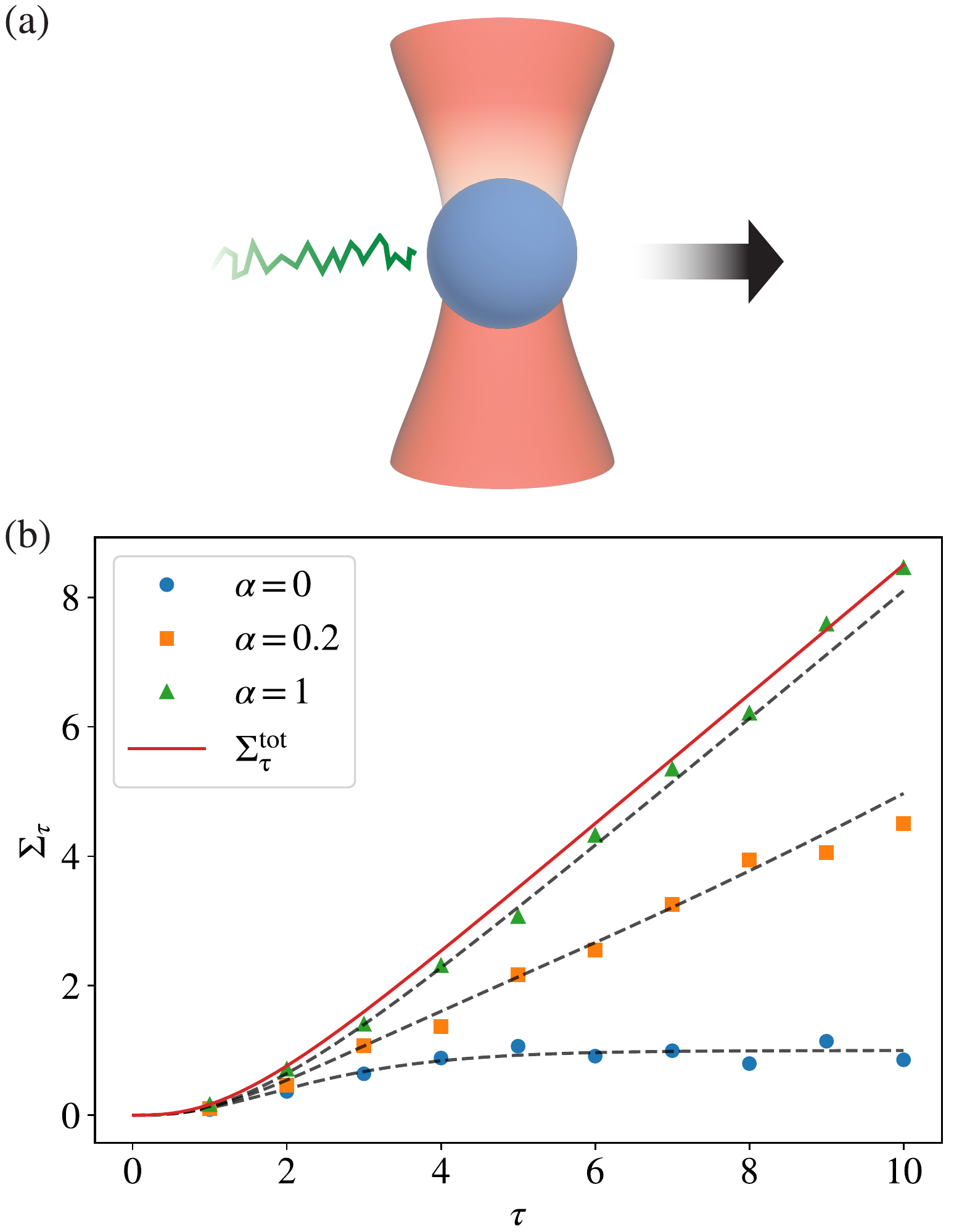}
    \caption{\label{fig:fig2} (a) Schematic illustration of the overdamped Brownian particle pulled by optical tweezers. 
    (b) Comparison between the true EP $\Sigma_\tau^{\rm tot}$ (solid line) and its lower bounds $\Sigma_\tau^{\rm b}$ according to the XTUR for various values of $\alpha$, the latter of which can be estimated empirically (symbols) as well as analytically (dashed lines). We choose $k = \gamma = T = v = 1$, and the empirical estimations are done using $10^6$ trajectories with $\Delta t = 0.01$ and $\Delta \omega = 0.5$.}
\end{figure}

Let us present a concrete example of the XTUR, which explicitly demonstrates the experimental testability of the relation. Towards this goal, we consider an overdamped Brownian particle pulled by optical tweezers, as shown in Fig.~\ref{fig:fig2}(a). Its motion is described by the Langevin equation
\begin{align}
    \gamma \dot{x} = -k(x - \omega v t) + \sqrt{2\gamma T}  \xi(t)
    \;,
\end{align}
where $\gamma$ is the friction coefficient, $T$ is the temperature of the bath, $k$ is the stiffness of the harmonic potential, $v$ is the speed of the optical tweezers, and $\omega$ is a parameter controlling the speed. We take the displacement of the particle as the current-like observable, which amounts to setting $\Lambda = 1$ in Eq.~\eqref{eq:general_observable_Langevin}:
\begin{align}
    \mathcal{J}_\tau ^\Lambda (\Gamma) = \int_0 ^\tau dt\, \dot{x} = x_\tau - x_0
    \;.
\end{align}
For the state-dependent observable whose general form is given in Eq.~\eqref{eq:general_observable_Langevin}, we use $g(x, t) = \frac{\alpha k}{\gamma} (x - \omega v t)$, where $\alpha$ controls the magnitude of the observable. This yields
\begin{align}
    \mathcal{K}_\tau ^g (\Gamma) = \frac{\alpha k}{\gamma} \int_0 ^\tau dt \,(x_t - \omega v t)
    \;.
\end{align}

Before proceeding further, some remarks on the role of $\alpha$ are in order. When $\alpha = 0$, the observable $\Theta_\tau^{\Lambda,g}$ is purely current-like, and the XTUR reduces to the ordinary TUR as discussed in Sec.~\ref{sec:XTUR_A}. Meanwhile, when $\alpha = 1$, $g$ has the form of $g^{\rm EB}$ shown in Eq.~\eqref{eq:gEB_expression}, in which case the XTUR becomes the EB as proved in Sec.~\ref{sec:XTUR2EB}. This clearly shows that the XTUR allows us to explore a much broader range of empirically accessible lower bounds on EP beyond the ordinary TUR and the EB.

Since this system has no odd-parity variables, all terms containing $s$ in Eq.~\eqref{eq:XTUR1} vanish. Thus, the XTUR reads
\begin{equation}
    \Sigma_\tau^{\rm b} \equiv \frac{2[(\tau \partial_\tau - \omega \partial_\omega) \langle \Theta_\tau^{\Lambda,g} \rangle - \langle \mathcal{K}_\tau^g \rangle ] ^2}{\text{Var}[\Theta_\tau^{\Lambda,g}]} \le  \Sigma_\tau^{\rm tot}  \;, \label{eq:XTUR_without_odd}
\end{equation}
where we use $\Sigma_\tau^{\rm b}$ to denote the lower bound of EP.

Note that $\mathcal{J}_\tau^\Lambda (\Gamma)$, $\mathcal{K}_\tau^g (\Gamma)$, and the composite observable $\Theta_\tau^{\Lambda,g} (\Gamma) = \mathcal{J}_\tau^\Lambda (\Gamma) + \mathcal{K}_\tau^g (\Gamma)$ are all experimentally measurable. Thus, to determine $\Sigma_\tau^{\rm b}$ experimentally, we only need to estimate the partial derivatives in its numerators, which can be done by the approximations
\begin{align}
    \partial_\tau \langle \Theta_\tau^{\Lambda,g} \rangle (\omega) &\approx \frac{\langle \Theta_{\tau+\Delta \tau}^{\Lambda,g} \rangle (\omega)- \langle \Theta_{\tau}^{\Lambda,g} \rangle (\omega)} {\Delta \tau} \;, \nonumber 
    \\  \partial_\omega \langle \Theta_\tau^{\Lambda,g} \rangle (\omega) &\approx \frac{\langle \Theta_\tau^{\Lambda,g} \rangle (\omega + \Delta \omega) - \langle \Theta_\tau^{\Lambda,g} \rangle (\omega)}{\Delta \omega}
\end{align}
by choosing sufficiently small $\Delta \tau$ and $\Delta \omega$.

Now, in Fig.~\ref{fig:fig2}(b), we compare $\Sigma_\tau^{\rm tot}$ (solid line) with the exact values of $\Sigma_\tau^{\rm b}$ (dashed lines) and their corresponding empirical estimates (symbols) as $\alpha$ is varied. For the case $\alpha = 0$, where only the current-like observable is taken into consideration, $\Sigma_\tau^{\rm b}$ falls far below $\Sigma_\tau^{\rm tot}$ as the duration $\tau$ is increased. In contrast, by choosing $\alpha = 1$, $\Sigma_\tau^{\rm b}$ can be made to stay very close to $\sigma_\tau^{\rm tot}$ for a broad range of $\tau$. As already discussed in Sec.~\ref{sec:XTUR_A}, the equality condition cannot be met in the TUR involving only current-like observables (which corresponds to $\alpha = 0$), while it is achievable in the XTUR using the composite observable. This example clearly demonstrates the importance of employing state-dependent observables to facilitate accurate estimation of EP.

\subsection{EB for two-level batteries} \label{sec:EB_ex}

\begin{figure}
    \includegraphics[width=\columnwidth]{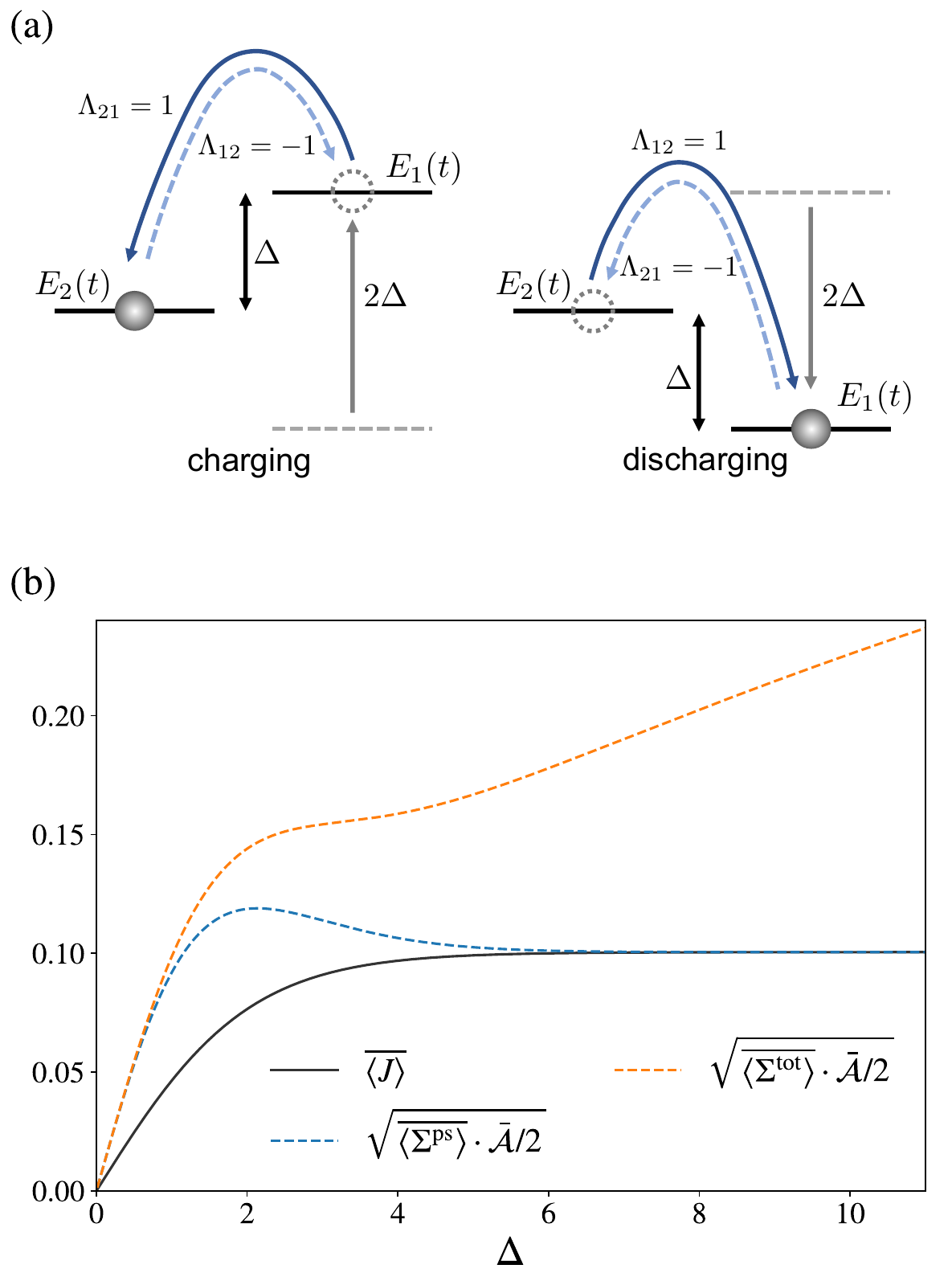}
    \caption{\label{fig:fig3} (a) Schematic illustration of the charging-discharging process. During the charging phase, the energy level of state~1 is increased, charging the battery by prompting the particle to jump to state~2. In contrast, during the discharging phase, the energy level of state 1 is decreased below that of state~2, resulting in battery discharge. In each phase, $\Delta$ represents the energy gap between the two states.
    (b) Verification of the EB~\eqref{eq:EB_in_two_state} for the charging-discharging process. The black solid line depicts the mean charging-discharging speed. The dashed curves represent the EBs set by the mean EP and the mean pseudo-EP.    
    }
    \end{figure}

As an example of the EB for Markov jump systems, we consider a charging-discharging process utilizing a two-level system, as depicted in Fig.~\ref{fig:fig3}(a). The energy of each state is denoted by $E_i (t)$ for $i=1,2$. During the charging process with duration $\tau_{\rm c}$, $E_1(t) > E_2(t)$ is maintained so that the particle tends to escape state $1$ to charge state $2$. Conversely, in the discharging process with duration $\tau_{\rm d}$, $E_1(t) < E_2(t)$ is maintained so that the particle is discharged from state $2$, releasing the energy $E_2(t) - E_1(t)$ via each transition. For simplicity, we assume that $E_2(t)$ is kept constant at $\Delta$, while $E_1(t) = 2\Delta$ during the charging process and $E_1(t) = 0$ during the discharging process. Then, keeping the local detailed balance, the transition rates are given by 
\begin{align} 
    R_{21} (t) &=  \frac{\mu(t)}{1+e^{-\beta [E_1(t)-\Delta]}} \,, 
    \nonumber\\ R_{12} (t) &=  \frac{\mu(t) e^{-\beta [E_1(t)-\Delta]}}{1+ e^{-\beta [E_1(t)-\Delta]}}
    \;,
\end{align}
where $\beta$ is the inverse temperature, and $\mu(t)$ controls the overall rate of transitions. Suppose we are concerned with quantifying the speed of the charging-discharging process. This can be quantified by the current-like observable $\mathcal{J}_\tau^{\Lambda}$ satisfying $\Lambda_{nm} (t) = \text{sgn}[E_m (t) - E_n (t)]$, which counts the excess number of charging and discharging transitions compared to the number of opposite transitions hindering the process.

Since $\Lambda_{nm}^2 = 1$ within this setup, $\chi_\tau$ in the EB~\eqref{eq:markov_EB} reduces to the dynamical activity $\mathcal{A}_\tau$. Then, the EB is rewritten as $\langle \mathcal{J}_\tau \rangle \le \sqrt{\mathcal{A}_\tau \Sigma _\tau/2}$. In the periodic state, we can define the mean charging-discharging speed as $\overline{\langle J \rangle} \equiv \langle \mathcal{J}_{\tau_{\rm c}+\tau_{\rm d}}^\Lambda \rangle / (\tau_{\rm c}+\tau_{\rm d})$, the mean EP as $\overline{\langle \Sigma^{\text{tot}}\rangle} \equiv \langle \Sigma_{\tau_{\rm c}+\tau_{\rm d}}^{\text{tot}}\rangle /(\tau_{\rm c}+\tau_{\rm d})$, the mean pseudo-EP as $\overline{\langle \Sigma^{\text{ps}}\rangle} \equiv \langle \Sigma_{\tau_{\rm c}+\tau_{\rm d}}^{\text{ps}}\rangle/(\tau_{\rm c}+\tau_{\rm d})$, and the mean dynamical activity as $\bar{\mathcal{A}} \equiv \mathcal{A}_{\tau_{\rm c}+\tau_{\rm d}}/(\tau_{\rm c}+\tau_{\rm d})$. Using these definitions, we obtain the upper bounds on the charging-discharging speed
\begin{align} \label{eq:EB_in_two_state}
    \overline{\langle J \rangle} \le \sqrt{\overline{\langle \Sigma^{\text{ps}}\rangle} \bar{\mathcal{A}} /2}  \le \sqrt{\overline{\langle \Sigma^{\text{tot}}\rangle} \bar{\mathcal{A}}/2 }
    \;.
\end{align}

The behaviors of $\overline{\langle J \rangle}$, $\sqrt{\overline{\langle \Sigma^{\text{ps}}\rangle} \bar{\mathcal{A}}/2 }$, and $\sqrt{\overline{\langle \Sigma^{\text{tot}}\rangle} \bar{\mathcal{A}}/2 }$ as functions of the energy gap $\Delta$ are shown in Fig.~\ref{fig:fig3}(b) for $\tau_{\rm c} = \tau_{\rm d} = 10$, $\mu(t) = 1$, and $\beta = 1$. Besides their consistency with the inequalities of Eq.~\eqref{eq:EB_in_two_state}, two points merit attention. First, the difference between the two upper bounds is negligible for small $\Delta$, which corresponds to the near-equilibrium regime. Second, the charging-discharging speed saturates to the tighter upper bound set by the pseudo-EP for large $\Delta$. This behavior stems from the equality condition of the tighter upper bound, which is given by
\begin{align} 
    \Lambda_{nm}^{\rm eq} = \frac{p_m R_{nm} - p_n R_{mn}}{p_m R_{nm} + p_n R_{mn}} 
    \;.
\end{align}
In the limit of large $\Delta$, since the transitions become almost totally irreversible, this quantity becomes either $1$ or $-1$ for every transition until the occupation probability of the higher energy level reaches zero. For transitions occurring afterwards, the quantity may have different values, but such transitions negligibly contribute to the physical quantities. Hence, the observable with $\Lambda_{nm} (t) = \Lambda_{nm}^{\rm eq}$ is practically equal to our chosen observable with $\Lambda_{nm} (t) = \text{sgn}[E_m (t) - E_n (t)]$, thus ensuring the saturation of the tighter bound. In contrast, the weaker bound by the total EP cannot be saturated in this regime, as the discrepancy between the EP and the pseudo-EP increases as the system moves farther away from equilibrium.

\begin{figure}
    \includegraphics[width=\columnwidth]{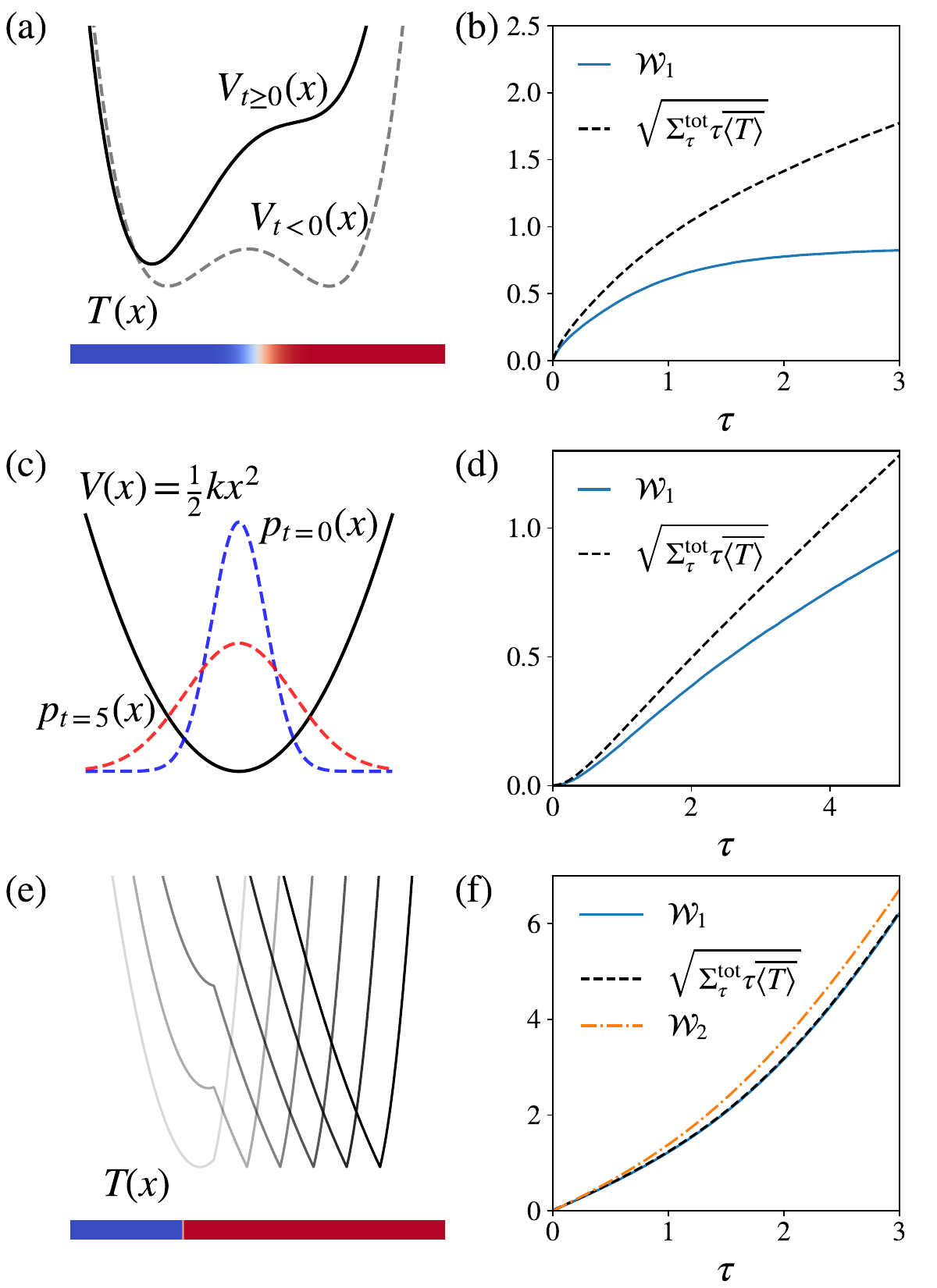}
    \caption{\label{fig:fig4} (a) Illustration of the tilted double-well potential and the nonuniform temperature field. The dashed and the solid lines represent the potential for $t<0$ and $t\geq 0$, respectively. The colormap below the potentials indicates the temperature field, which is colder (hotter) on the left (right).
    (b) Verification of the CSL~\eqref{eq:CSL_Wasserstein} for the process depicted in (a).
    (c) Illustration of the probability distribution (dashed curves) broadening within a harmonic potential (solid curve) as temperature increases with time.
    (d) Verification of the CSL~\eqref{eq:CSL_Wasserstein} for the process depicted in (c).
    (e) Optimal protocol $V_{\text{EQ}} (x,t)$ that ensures the equality condition of the CSL~\eqref{eq:CSL_Wasserstein}. Lighter (darker) curves correspond to the potentials at earlier (later) times, and the colormap below the potentials indicates the temperature field.    
    (f) Verifications of the saturation of the CSL for the Wasserstein-$1$ distance and its violation for the Wasserstein-2 distance. }
\end{figure}

\subsection{CSL in Langevin systems subjected to inhomogeneous temperature field} \label{ex_CSL_Langevin}

To illustrate the CSL in a nonuniform temperature field, we consider a one-dimensional Brownian particle trapped in either a single-well or a double-well potential~\cite{Lee2022}. Its equation of motion is given by 
\begin{align} \label{eq:inhomo_temp_eom}
    \gamma \dot{x} = - \frac{\partial V(x,t)}{\partial x} + \sqrt{2 \gamma T(x,t)}  \bullet \xi (t)
    \;,
\end{align}
where $\xi(t)$ is a Gaussian white noise satisfying $\langle \xi(t)\rangle = 0$ and $\langle \xi(t)\xi(t') \rangle = \delta(t-t')$. We note that the spatial inhomogeneity in the noise amplitude stems only from the temperature field $T(x,t)$ and not from the constant friction coefficient $\gamma$, in which case the It\^o description of the multiplicative noise is indeed appropriate in the overdamped limit, as discussed in Ref.~\cite{Durang2015}. 

We illustrate the CSL for three different processes of the system. The first example concerns the relaxation of the Brownian particle when the potential $V(x)$ and the temperature field $T(x)$ are abruptly changed. More specifically, we consider the potential
\begin{align} 
    V(x) &= V_{\text{DW}}(x) + \Theta_{\text{tilt}}\frac{x}{x_m} \,, 
\end{align}
where $ V_{\text{DW}}(x) = E_b [ \left( x/x_m \right)^4 - 2\left( x/x_m \right)^2 ] $ is a double-well potential, and $\Theta_{\text{tilt}}$ controls the magnitude of the linear tilting potential. This potential has widely been used to investigate the finite-time Landauer principle~\cite{Proesmans2020, Lee2022} governing the entropic cost of erasing one bit of information. Meanwhile, the temperature field is given by
\begin{align} 
    T(x) &= T_0 \left( 1+\delta \tanh \frac{x}{x_T} \right)
    \,,
\end{align}
where $\delta$ and $x_T$ control the magnitude and the length scale of temperature change across the system, respectively.
For $t < 0$, both $\Theta_\textrm{tilt}$ and $\delta$ are set to zero, letting the particle equilibrate in the bare double-well potential $V_{\text{DW}}(x)$ to the uniform temperature field $T(x) = T_0 = 1$. Then, at $t = 0$, we set $\Theta_{\rm tilt}=2$ and $\delta = 0.5$, switching on the tilting potential and the nonuniform temperature field, as illustrated in Fig.~\ref{fig:fig4}(a). This protocol amounts to erasing one bit of information encoded by the Brownian particle by tuning both the potential and the temperature field. Given the other parameters fixed at $\gamma = 1 $, $x_m = 1 $, $x_T = 0.2$, and $E_b = 1$, Fig.~\ref{fig:fig4}(b) verifies the CSL stated in Eq.~\eqref{eq:CSL_Wasserstein}, where $\overline{\langle T \rangle} \equiv \frac{1}{\tau} \int_0 ^\tau dt \langle T(x, t)\rangle$ plays the role of $\overline{\langle \rho (\mathbb{D}) \rangle}$. Thus, Eq.~\eqref{eq:CSL_Wasserstein} provides a way to study the finite-time Landauer principle involving the local heating as well as the potential manipulation.

The next example pertains to the Brownian particle confined in a harmonic potential $V(x) = \frac{1}{2} k x^2$ with the temperature of the heat bath changing in time as
\begin{align} \label{eq:ex2_protocol}
    T(t) = \begin{cases}
    T_0 & \text{for $t < 0$}\;,\\
 	T_0 (1+t/\tau_{\rm T}) &\text{for $t \ge 0$}\;.
 \end{cases}
\end{align}
This means that the particle distribution becomes broader as time goes by, as illustrated in Fig.~\ref{fig:fig4}(c). Choosing $k=1$ and $\tau_{\rm T} = 1/4$, in Fig.~\ref{fig:fig4}(d) we demonstrate the validity of the CSL stated in Eq.~\eqref{eq:CSL_Wasserstein}.

The final example demonstrates the violation of the CSL when $\mathcal{W}_2$ is substituted for $\mathcal{W}_1$ in Eq.~\eqref{eq:CSL_Wasserstein}. In this example, the position-dependent temperature field is given by
\begin{align}
    T(x) = \begin{cases} T_1 & \text{for } x<0
    \\ T_2 & \text{for } x\ge0
    \end{cases}
    \;.
\end{align}
In the initial state, the Brownian particle has the equilibrium distribution $p(x,t=0)$ corresponding to $T_1 = T_2 = 1$, $\gamma=1$, and $V(x) = \frac{1}{2} k (x-x_0)^2$ with $k=1$ and $x_0 = -2$. When $t \geq 0$, the temperature field is abruptly changed to $T_1 = 1$ and $T_2 = 4$, and the particle is dragged by the potential $V_{\text{EQ}}(x,t)$ [see Fig.~\ref{fig:fig4}(e)], which is chosen to be the protocol ensuring the equality condition of the CSL~\eqref{eq:CSL_Wasserstein}. See Appendix~\ref{Appendix_c} for further elaborations on the equality condition of the CSL~\eqref{eq:CSL_Wasserstein} and the explicit form of $V_{\text{EQ}}(x,t)$. For this system, we numerically obtained the probability distribution function $p(x,t)$, from which $\mathcal{W}_1$, $\mathcal{W}_2$, and $\Sigma_\tau ^{\text{tot}}$ can be calculated.

In one-dimensional systems, the Wasserstein-$k$ distance is given by~\cite{Santambrogio2015}
\begin{equation}
\mathcal{W}_k (p,q) = \left( \int_0^1 dr \left|F_p^{-1}(r) - F_q^{-1}(r) \right|^{k} \right)^{1/k},
\end{equation}
where $F_p(x) \equiv \int_{-\infty}^x dx^\prime p(x^\prime, t)$ is the cumulative distribution function of $p$, and $F_p^{-1}(r)$ is its inverse function. Thus, except for the case where the particle distribution simply translates while maintaining its shape as $p(x,t) = p(x-ct)$, the initial and the final distributions always satisfy $\mathcal{W}_2 > \mathcal{W}_1$. Note that our protocol saturates the CSL~\eqref{eq:CSL_Wasserstein}, and $p(x,t)$ changes its shape as the system evolves. Hence, the substitution of $\mathcal{W}_2$ for $\mathcal{W}_1$ violates the CSL, that is, 
\begin{align}
    \Sigma_\tau ^{\text{tot}} \tau \overline{\langle T \rangle} = \mathcal{W}_1 ^2 < \mathcal{W}_2 ^2
\end{align}
as displayed in Fig.~\ref{fig:fig4}(f).


\section{Conclusion} \label{sec:conclusions}

\begin{table}[] 
\caption{The table lists the sections where one can find the derivation of inequalities discovered through the hierarchical structure. For already known relations, the relevant references are marked. "OL" and "UL" stand for overdamped Langevin and underdamped Langevin dynamics, respectively.}
\begin{center}
\begin{threeparttable}
\begin{tabular}{@{}cccc@{}}
\midrule \midrule
\multicolumn{1}{c}{\multirow{2}{*}{Inequality}} & \multicolumn{3}{c}{Type of dynamics}                                     \\ \cmidrule(l){2-4} 
\multicolumn{1}{c}{}    & Markov jump & \qquad OL \qquad \qquad & \qquad UL \qquad \qquad \\ \cmidrule(l){1-4}
XTUR        & Sec.~\ref{sec:XTUR} & Sec.~\ref{sec:XTUR} & Sec.~\ref{sec:XTUR} \\
TUR         & ~\cite{Koyuk2020} & ~\cite{Koyuk2020} & Sec.~\ref{sec:XTUR} \\
EB          & Sec.~\ref{sec:XTUR2EB} & ~\cite{Dechant2018B} & ~\cite{Dechant2018B} \\
CSL         & ~\cite{Dechant2022,Lee2022,VanVu2023} & Sec.~\ref{sec:EB2CSL} & N/A\tnote{1}  \\
PE tradeoff & Sec.~\ref{sec:EB2PE} & ~\cite{Dechant2018B} & ~\cite{Dechant2018B}\ \\ 
\midrule \midrule
\end{tabular} 
\begin{tablenotes}
\item[1] The expression is not available. \\

\end{tablenotes}
\end{threeparttable}
\end{center}
\label{table:whatsnew}
\end{table}

Through this study, we have unveiled the hierarchical connections among various thermodynamic tradeoff relations, encompassing the TUR, the EB, the CSL, and the PE tradeoff. This hierarchy commences with the XTUR, which deals with the most general types of observables. All the other tradeoff relations stem from the XTUR by imposing specific constraints upon the associated observable. When state-dependent observables are excluded, the XTUR simplifies to the TUR. On the other hand, choosing state-dependent observables as described in Eqs.~\eqref{eq:gEB_expression} and \eqref{eq:gEB_expression_Markov} transforms the XTUR into the EB. Further down the hierarchy, focusing on the systems without odd-parity variables, the CSL and the PE tradeoff follow from the EB by employing the heat current and the IPM as the associated current-like observable, respectively. 

While completing the hierarchy, we have established various tradeoff relations, which are summarized in the following as well as in Table~\ref{table:whatsnew}. First, we derived the EB and the PE tradeoff for the Markov jump process driven by general time-dependent protocols. To our knowledge, the former was never discussed under any circumstances in the literature, while the latter was known only for the steady-state engines. We also explicitly demonstrated the EB for the two-level system in Sec.~\ref{sec:EB_ex}. Second, we discovered an expression of the CSL applicable to overdamped Langevin systems whose diffusion matrix in nonuniform in space and time. The expression greatly extends the applicability of the CSL, which have been limited to systems with the uniform, time-independent diffusion matrix. We numerically verified the CSL in Sec.~\ref{ex_CSL_Langevin} for one-dimensional examples with inhomogeneous temperature fields. Finally, we found an expression of the TUR for underdamped Langevin dynamics, which fits into the entire hierarchical relationship. We stress that the discovery of all these relations have been motivated and facilitated by our unified hierarchy. 

Furthermore, this hierarchy offers a fundamental and comprehensive perspective on the landscape of significant thermodynamic tradeoff relations that have emerged in the field of nonequilibrium thermodynamics over the past decade. In particular, the structure indicates that the TUR, the EB, the CSL, and the PE tradeoff can all be categorized under a single class of thermodynamic relations, which we might call the {\em XTUR class}.
Hence, if future research uncovers another thermodynamic inequality, one critical task will be to ascertain whether this new relation falls within the XTUR class or deviates from it. The latter case would suggest that the inequality is truly {\em new} and may open up the avenue towards a hitherto unexplored realm of thermodynamics.


For the future research, it is highly desirable to extend our theory to open quantum systems~\cite{Horowitz2012, Horowitz2013, Manzano2015, Manzano2018, Shiraishi2019, Funo2019, Hasegawa2020, Hasegawa2021, VanVu2021, VanVu2022, VanVu2023}. Over the past decade, numerous thermodynamic tradeoff relations have also emerged in the quantum domain, such as the quantum TURs~\cite{Hasegawa2020,Hasegawa2021,VanVu2022,Hasegawa2023} and the quantum speed limits~\cite{Funo2019, VanVu2021, VanVu2023, Hasegawa2023}. However, these relations have largely been investigated as independent properties. Therefore, a critical forthcoming task is to establish a unified hierarchical framework for these tradeoff relations governing open quantum systems. This will facilitate a deeper understanding of quantum thermodynamics and its applications.

{\em Acknowledgments.} --- The authors thank Su-Chan Park for fruitful discussions. This research has been supported by the POSCO Science Fellowship of POSCO TJ Park Foundation (E.K. and Y.B.), KIAS Individual Grant No. PG064901 (J.S.L.), and an appointment to the JRG Program at the APCTP through the Science and Technology Promotion Fund and Lottery Fund of the Korean Government (J.-M.P.). This was also supported by the Korean Local Governments - Gyeongsangbuk-do Province and Pohang City (J.-M.P.).

\begin{appendix}

\section{Detailed derivations}

\subsection{Fisher information in Langevin systems, Eq.~\eqref{eq:FI_at_0}} \label{secA:eval_Fisher}

When the system follows the modified Langevin dynamics governed by Eq.~\eqref{eq:modified_drift}, its path probability satisfies~\cite{Onsager1953}
\begin{align}  \label{eqA:path_prob_theta}
    &\mathcal{P}_\theta [\Gamma] = \mathcal{N} p(0)\,e^{-\frac{1}{4} \int_0 ^\tau dt\, (\dot{\mathbf{x}} - \mathbf{A}_\theta)^{\textsf T} \bullet \mathbb{D}^{-1} \bullet (\dot{\mathbf{x}} - \mathbf{A}_\theta)},
\end{align}
where $\mathcal{N}$ is the normalization constant, and $p(0) \equiv p(\textbf{x},0;s,\omega)$ is the initial probability distribution. 
Then, we have
\begin{align} 
    & \partial^2 _\theta \ln \mathcal{P}_\theta = \frac{1}{2} \int_0 ^\tau dt \, \partial^2 _\theta \mathbf{A}_\theta^{\textsf T} \mathbb{D}^{-1} \bullet (\dot{\mathbf{x}}-\mathbf{A}_\theta)
    \nonumber\\ & \,\,\,\,\,\,\,\, - \frac{1}{2} \int_0 ^\tau dt \, \partial_\theta \mathbf{A}_\theta^{\textsf T} \mathbb{D}^{-1} \partial_\theta \mathbf{A}_\theta
     + \frac{\partial_\theta^2 p(0)}{p(0)}- \frac{\{\partial_\theta p(0)\}^2}{p(0)^2}.
\end{align}
Using $\partial_\theta \mathbf{A} _\theta \big|_{\theta =0} = \mathbf{J}^{\text{irr}}/p$ and $\langle \dot{\textbf{x}}-\textbf{A}_\theta\rangle_\theta = 0$, the Fisher information at $\theta =0$ is obtained as
\begin{align} 
    \mathcal{I}(0) 
    &=  \langle -\partial^2 _\theta \ln \mathcal{P}_\theta \rangle_{\theta = 0}
    \nonumber\\ 
    &=\frac{1}{2} \int_0 ^\tau dt \int d\mathbf{x}\, \frac{{\mathbf{J}^\text{irr}}^{\textsf T} \mathbb{D}^{-1} \mathbf{J}^{\text{irr}}}{p}\nonumber\\
    &\qquad +\left\langle  \left[ (s\partial_s + \omega \partial_\omega)\ln p(0) \right]^2 \right\rangle 
    = \frac{1}{2} \left( \Sigma^{\text{tot}}_\tau +\mathcal B \right)
    \;,
\end{align}
where $ \mathcal{B} = 2\langle  \left[ (s\partial_s + \omega \partial_\omega)\ln p(0) \right]^2 \rangle$. If we choose an initial distribution independent of $s$ and $\omega$, then $\mathcal{B} = 0$. 

\subsection{Evaluation of $\partial_\theta \langle \mathcal J_\tau^\Lambda \rangle |_{\theta=0} $ and $\partial_\theta \langle \mathcal K_\tau^g \rangle |_{\theta=0} $, Eq.~\eqref{eq:JK_calc} } \label{secA:evalu_J_K}

We begin by evaluating $\partial_\theta \langle \mathcal J_\tau^\Lambda \rangle |_{\theta=0} $. Through direct substitution, we can show that the total probability current $\mathbf{J}_\theta \equiv \mathbf{J}_\theta^{\rm rev} +(1+\theta)\mathbf{J}_\theta^{\rm irr}$ of the modified dynamics governed by Eq.~\eqref{eq:modified_drift} is related to the original current by $\mathbf{J}_\theta ({\textbf x},t;s,\omega) = (1+\theta) \mathbf{J} ({\textbf x},t_\theta;s_\theta,\omega_\theta)$. Using this relation, the mean of $\mathcal{J}_\tau^\Lambda$ in the modified dynamics is obtained as
\begin{align} 
    \langle \mathcal{J}_\tau^\Lambda \rangle_{\theta} & = \int_0 ^\tau dt \int d\mathbf{x}\, \boldsymbol\Lambda^{\textsf T} ({\textbf x},\omega t;s)  \mathbf{J}_{\theta} ({\textbf x},t;s,\omega)
    \nonumber\\ 
    & =  \int_0 ^\tau dt \int d\mathbf{x}\,  (1+\theta)\boldsymbol\Lambda^{\textsf T}({\textbf x},\omega t;s)  \mathbf{J} ({\textbf x},t_\theta; s_\theta, \omega_\theta)
    \nonumber\\ 
    &= \int_0 ^{\tau_\theta} dt \int d\mathbf{x}\, \boldsymbol\Lambda^{\textsf T} \left({\textbf x},\omega_\theta t;s\right) \mathbf{J}({\textbf x},t;s_\theta,\omega_\theta)
    \;.
\end{align}
The Taylor expansion $\boldsymbol{\Lambda} (\mathbf{x}, \omega_\theta t;s) = \boldsymbol{\Lambda} (\mathbf{x}, \omega_\theta t;s_\theta) + \theta s_\theta \partial_s \boldsymbol{\Lambda} (\mathbf{x}, \omega_\theta t;s) |_{s=s_\theta}  + O(\theta^2)$ yields
\begin{align} \label{eqA:mean_J_theta}
    \langle \mathcal{J}_\tau^\Lambda \rangle_{\theta}
    &=  \langle \mathcal{J}_{\tau_\theta}^\Lambda \rangle |_{s=s_\theta, \omega=\omega_\theta} \nonumber \\
    &+ \theta  \int_0^{\tau_\theta} dt \, s_\theta \partial_s \boldsymbol{\Lambda}^{\textsf T} |_{s=s_\theta} \mathbf{J} ({\textbf x},t;s_\theta,\omega_\theta) + O ( \theta^2 ).
\end{align}
Differentiating both sides of this equation with respect to $\theta$ and setting $\theta=0$, we arrive at
\begin{align} 
    \partial_\theta \langle \mathcal J_\tau^\Lambda  \rangle _\theta \big|_{\theta=0} &= \hat O_\tau \langle \mathcal J_\tau^\Lambda \rangle +\int_0^\tau dt~ \langle (s\partial_s \boldsymbol{\Lambda}^{\textsf T} ) \circ \dot {\textbf x} \rangle   \,.
\end{align}

Now, we turn to the evaluation of $\partial_\theta \langle \mathcal K_\tau^g \rangle |_{\theta=0} $. The mean of $\mathcal{K}_\tau^g$ in the modified dynamics satisfies
\begin{align}
    \langle \mathcal{K}_\tau^g \rangle _\theta
    &= \int_0^\tau dt \int d\mathbf{x}\, g(\mathbf{x}, \omega t;s) p_\theta(\mathbf{x},t;s,\omega) \nonumber \\
    &= \frac{1}{1 + \theta} \int_0^{\tau_\theta} dt \int d\mathbf{x}\, g(\omega_\theta t;s) p(t;s_\theta,\omega_\theta)\,.
\end{align}
Proceeding with the Taylor expansion $g(\mathbf{x}, \omega_\theta t;s) = g(\mathbf{x}, \omega_\theta t;s_\theta) + \theta s_\theta \partial_s g(\mathbf{x}, \omega_\theta t;s) |_{s=s_\theta}  + O(\theta^2)$, we obtain
\begin{align}
    \langle \mathcal{K}_\tau^g \rangle _\theta &= \frac{1}{1 + \theta}  \langle \mathcal{K}_{\tau_\theta} \rangle |_{s=s_\theta,\omega = \omega_\theta} \nonumber \\
    &+ \theta  \int_0^\tau dt\, s_\theta \partial_s g |_{s=s_\theta}  p(t;s_\theta,\omega_\theta) + O (\theta^2) \,.
\end{align}
Differentiating both sides with respect to $\theta$, we finally derive
\begin{align} 
    \partial_\theta \langle \mathcal K_\tau^g  \rangle _\theta \big|_{\theta=0} &= \hat O_\tau \langle \mathcal K_\tau^g \rangle - \langle \mathcal K_\tau^g \rangle + \int_0^\tau dt~ \langle s\partial_s g \rangle \;.
\end{align}

\subsection{XTUR saturation for Langevin systems, Eq.~\eqref{eq:equa_cond_Langevin} } \label{Appendix_Equa_cond_Langevin}

From the definition of $\Theta_\tau^{\Lambda,g}$, the left-hand side of Eq.~\eqref{eq:eq_condition} at $\theta =0$ can be written as
\begin{align} \label{eqA:LHS1}
    \Theta_\tau^{\Lambda,g} (\Gamma) - \langle \Theta _\tau^{\Lambda,g} \rangle = &\int_0^\tau dt \left( \boldsymbol{\Lambda}^{\textsf T} \circ \dot{{\textbf x}} + g   \right) \nonumber \\
    &- \int_0^\tau dt\int d{\textbf x} \left( \boldsymbol{\Lambda}^{\textsf T} {\textbf J} + g p \right) \,.  
\end{align}

To compute the right-hand side of Eq.~\eqref{eq:eq_condition}, we must use the conversion relationship between the Stratonovich and the It\^o products
\begin{equation} \label{eqA:Ito_Strato}
\mathbf{h}^{\textsf T}(\mathbf{x},t)\circ d\mathbf{x} = \mathbf{h}^{\textsf T}(\mathbf{x},t) \bullet d\mathbf{x} + \text{tr}[\mathbb{D}\nabla \mathbf{h}^{\textsf T}(\mathbf{x},t)]\,dt \;,
\end{equation}
which holds for an arbitrary function $\mathbf{h}(\mathbf{x},t)$. Here, we have used the relation $d\mathbf{x} = \mathbf{A} dt + \mathbb{B} \bullet d\mathbf{W}$, where $\mathbf{W}$ stands for the Wiener process. To derive this formula, one starts from the definition of the Stratonovich product
\begin{align} \label{eqA:Strato_expansion}
    h_i(\mathbf{x} ,t) \circ dx_i = \frac{1}{2}\left[ h_i(\mathbf{x} + d\mathbf{x},t+dt) + h_i(\mathbf{x},t) \right] dx_i \;.
\end{align}
Keeping the terms up to the order of $dt$, the right-hand side can be expanded as
\begin{align} \label{eqA:Strato_expansion1}
    h_i(\mathbf{x} ,t) \circ dx_i = h_i (\mathbf{x} ,t) \bullet dx_i + \frac{1}{2} \sum_j \partial_{x_j} h_i(\mathbf{x} ,t) dx_i dx_j \;.
\end{align}
Using $dx_i dx_j = \sum_{k,l} \mathbb{B}_{ik} \mathbb{B}_{jl} dW_k dW_l + O(dt^{3/2})$ and the heuristic It\^o rule $dW_k dW_l = \delta_{kl} dt$, this simplifies to
\begin{align} \label{eqA:Strato_expansion2}
    h_i(\mathbf{x} ,t) \circ dx_i = h_i (\mathbf{x} ,t) \bullet dx_i + \sum_{j}  \mathbb{D}_{ij}\partial_{x_j} h_i(\mathbf{x} ,t)dt \;.
\end{align}
Summing this equation side by side over $i$, we derive the conversion formula shown in Eq.~\eqref{eqA:Ito_Strato}. 

Using Eqs.~\eqref{eqA:path_prob_theta}, ~\eqref{eqA:Ito_Strato}, and \eqref{eq:modified_drift}, we can rewrite Eq.~\eqref{eq:eq_condition} at $\theta =0$ as 
\begin{align} \label{eqA:RHS1}
    &\partial_\theta \log \mathcal{P}_\theta (\Gamma) \big|_{\theta = 0} \nonumber \\
    &= \frac{1}{2} \int_0 ^\tau dt~ \frac{{\mathbf{J}^{\text{irr}}}^{\textsf T}}{p} \mathbb{D}^{-1} \bullet (\dot{\mathbf{x}} - \mathbf{A})
    \nonumber\\ &= \frac{1}{2} \int_0 ^\tau dt~ \frac{{\mathbf{J}^{\text{irr}}}^{\textsf T}}{p} \mathbb{D}^{-1} \circ \dot{\mathbf{x}} - \frac{1}{2} \int_0 ^\tau dt~ \text{tr}\left[\mathbb{D}\nabla\left(\frac{{\mathbf{J}^{\text{irr}}}^{\textsf T}}{p} \mathbb{D}^{-1} \right) \right]
    \nonumber\\ 
    & ~~~- \frac{1}{2} \int_0 ^\tau dt \left[  \frac{{\mathbf{J}^\text{irr}}^{\textsf T} \mathbb{D}^{-1} \{\mathbf{J}+ (\nabla^{\textsf T} \mathbb{D}p)^{\textsf T} )\}}{p^2} \right]
    \;.
\end{align}
Here we note that
\begin{align} \label{eqA:note_trace} &\textrm{tr}\left[\mathbb{D}\nabla\left(\frac{{\mathbf{J}^{\text{irr}}}^{\textsf T}}{p} \mathbb{D}^{-1} \right) \right] = \mathbb{D}_{ij} \partial_{x_j} \!\left[\frac{J^{\text{irr}}_k (\mathbb{D}^{-1})_{ki}}{p} \right]
    \nonumber\\ & = \left[\partial_{x_j} (\mathbb{D}^{-1})_{ki} \right] \mathbb{D}_{ij} \frac{J^{\text{irr}}_k}{p} + \nabla^{\textsf T} \left(\frac{ \mathbf{J}^{\text{irr}}}{p}\right)
    \nonumber\\ & = -\frac{{\mathbf{J}^\text{irr}}^{\textsf T}\mathbb{D}^{-1}}{p} (\nabla^{\textsf T} \mathbb{D})^{\textsf T}  - \frac{\nabla p^{\textsf T} \mathbf{J}^{\text{irr}}}{p^2} + \frac{\nabla^{\textsf T}  \mathbf{J}^{\text{irr}}}{p}
    \nonumber\\ & = -\frac{{\mathbf{J}^\text{irr}}^{\textsf T} \mathbb{D}^{-1}(\nabla^{\textsf T} \mathbb{D}p)^{\textsf T }}{p^2} + \frac{\nabla^{\textsf T}  \mathbf{J}^{\text{irr}}}{p}
    \;.
\end{align}
For the third equality in Eq.~\eqref{eqA:note_trace}, $\mathbb{D}_{ij} \partial_{x_j} (\mathbb{D}^{-1})_{ki} = - (\mathbb{D}^{-1})_{ki} \partial_{x_j} \mathbb{D}_{ij} $ has been used since $\partial_{x_j}[(\mathbb{D}^{-1})_{ki}\mathbb{D}_{ij}] =\partial_{x_j} \delta_{kj} =0 $. By substituting the outcome of Eq.~\eqref{eqA:note_trace} into Eq.~\eqref{eqA:RHS1}, we arrive at 
\begin{align}  \label{eqA:RHS2}
    &\partial_\theta \log \mathcal{P}_\theta (\Gamma) \big|_{\theta = 0} \nonumber \\
    &=\frac{1}{2} \int_0 ^\tau dt \left(\frac{{\mathbf{J}^\text{irr}}^{\textsf T}}{p} \mathbb{D}^{-1} \circ \dot{\mathbf{x}} 
    -\frac{ {\mathbf{J}^\text{irr}}^{\textsf T} \mathbb{D}^{-1} \mathbf{J}}{p^2} - \frac{\nabla^{\textsf T}  \mathbf{J}^{\text{irr}}}{p} \right)
    \;.
\end{align}
Finally, plugging Eqs.~\eqref{eqA:LHS1} and \eqref{eqA:RHS2} into Eq.~\eqref{eq:eq_condition}, we obtain the equality condition stated in Eq.~\eqref{eq:equa_cond_Langevin}.

\subsection{Fisher information in Markov jump systems, Eq.~\eqref{eq:FI_at_0_markov} }
\label{sec:eval_Fisher_Markov}

When the system follows the modified Markov jump process satisfying Eq.~\eqref{eq:perturbed_master_eq}, then its path probability is given by~\cite{Esposito2010, Seifert2012}
\begin{align} \label{eq:path_markov}
    \mathcal{P}[\Gamma] &= p_{n_0} (0) \exp \bigg[ \int_0 ^\tau dt \sum_n \eta_n (t) R_{nn}^\theta (\omega t)
    \nonumber\\ 
    & ~~~~~~~~~~~~~ + \int_0 ^\tau dt \sum_{n \neq m} \ln R^\theta _{nm} (\omega t)  \dot{N}_{nm} \bigg]
    \;,
\end{align}
where $p_{n_0} (0)$ represents the initial-state distribution, set to be independent of the perturbing parameters. Note that the first and the second terms in the exponent correspond to the sums of the staying and the transition probabilities, respectively. Then, the Fisher information satisfies
\begin{align} 
    &\mathcal{I}(\theta) = - \langle \partial ^2 _\theta \log \mathcal{P}[\Gamma] \rangle_\theta
    \nonumber\\ 
    &= - \int_0 ^\tau dt  \sum_n p^\theta _n \partial ^2 _\theta R^\theta _{nn} -  \int_0 ^\tau dt \sum_{n \neq m} p^\theta _m R^\theta _{nm} \partial ^2 \ln R^\theta _{nm} 
    \nonumber\\ 
    &= \int_0 ^\tau dt \sum_{n \neq m} \left( p_m^\theta \partial^2 _\theta R^\theta _{nm} - p^\theta _m R^\theta _{nm} \partial^2 _\theta \ln R^\theta _{nm} \right)
    \nonumber\\ 
    &= \int_0 ^\tau dt \sum_{n \neq m} p^\theta _m R^\theta _{nm} (\partial _\theta \ln R^\theta _{nm})^2
    \;.
\end{align}
Hence, the Fisher information at $\theta = 0$ is obtained as
\begin{align} 
    \mathcal{I}(0) &= \int_0 ^\tau dt \sum_{n \neq m} p_m R_{nm} (K_{nm}^{\theta=0})^2 \nonumber\\
    &= \frac{1}{2} \int_0^\tau dt \sum_{n\neq m} a_{nm} (K_{nm}^{\theta=0})^2 = \frac{1}{2} \Sigma_\tau^\text{ps} 
    \;.
\end{align}


\subsection{Variance of the current-like observable\\in Markov jump processes, Eq.~\eqref{eq:Var_EB} } \label{Appendix_A4.5}

The number of jumps from state~$m$ to state~$n$ during an infinitesimal time interval $dt$, namely $\dot{N}_{nm}dt = N_{nm}(t+dt)-N_{nm}(t)$, can be factorized as $\eta_{m}(t) T_{nm}(t)dt$, where $T_{nm}(t)dt$ denotes the number of jump from state $m$ to state $n$ {\em provided that} the system is in the state $m$ at time $t$. Here, $T_{nm}(t)dt = 1$ with probability $R_{nm}(t)dt$ and $T_{nm}dt = 0$ with probability $1- R_{nm}(t)dt$, which implies $\langle T_{nm}(t)dt \rangle = \langle [T_{nm}(t)dt]^2 \rangle  = R_{nm}(t)dt $. Since the jumps are independent of each other, we have
\begin{align} \label{eqA:variance_Tnm}
    &\langle [T_{nm}(t)-R_{nm}(t)][T_{n'm'}(t')-R_{n'm'}(t')] \rangle \nonumber \\
    &= R_{nm}(t) \delta_{n n'} \delta_{m m'}\delta(t-t').
\end{align}
Therefore, using Eq.~\eqref{eq:EB_Markov_observable}, the variance of the composite observable $\Theta_\tau^{\Lambda,g^{\rm EB}}$ is calculated as
\begin{align} \label{eqA:Variance_evaluation1}
    &\text{Var}(\Theta _\tau^{\Lambda,g^{\rm EB}}) = \langle (\Theta_\tau^{\Lambda,g^{\rm EB}})^2 \rangle \nonumber \\
    &= \int_0^\tau dt \int_0^\tau dt' \sum_{n \neq m} \sum_{n' \neq m'} \Lambda_{nm}(\omega t) \Lambda_{n'm'}(\omega t')
    \nonumber \\
    & \times \langle \eta_{m}(t)\eta_{m'}(t')[T_{nm}(t)-R_{nm}(t)][T_{n'm'}(t')-R_{n'm'}(t')] \rangle.
\end{align}
Note that $\eta_m(t)$ and $T_{nm}(t)$ are independent of each other, allowing the factorization of the ensemble average to $\langle \eta_{m}(t)\eta_{m'}(t') \rangle \langle [T_{nm}(t)-R_{nm}(t)][T_{n'm'}(t')-R_{n'm'}(t')] \rangle$. Therefore, using Eq.\eqref{eqA:variance_Tnm}, we obtain
\begin{align}
    \text{Var}(\Theta _\tau^{\Lambda,g^{\rm EB}}) 
    &= \int_0^\tau dt \sum_{n \neq m} \Lambda_{nm}(\omega t)^2 \langle \eta_{m}(t)^2 \rangle R_{nm}(\omega t) \nonumber \\
    &= \int_0^\tau dt \sum_{n \neq m} \Lambda_{nm}(\omega t)^2 p_m (t) R_{nm}(\omega t). 
\end{align}

\subsection{Wasserstein-1 distance on graphs\\ and derivation of Eq.~\eqref{eq:Lip_1_other_expression} } \label{Appendix_A5}

We start with a brief overview of the theory of optimal transport on graphs~\cite{VanVu2023}. Let $\mathcal{G}(V,E)$ represent an undirected graph, where $V$ denotes the set of all states, and $E$ the set of all state pairs joined by allowed transitions. For any given path $P$, $\text{len}(P)$ denotes its length, defined as the number of edges belonging to the path. Using this notation, the shortest-path distance between states $n$ and $m$, denoted as $d_{nm}$, can be expressed as $d_{nm} \equiv \min_P { \text{len}(P) }$. Notably, $d_{nm}$ serves as a metric applicable to the graph $\mathcal{G}(V,E)$. In this context, the Wasserstein-1 distance between two distributions, $p$ and $q$, on the graph $\mathcal{G}(V,E)$ is defined as
\begin{align} 
    \mathcal{W}_1 (p, q) \equiv \min _{\pi_{nm} \in \Phi (p, q)} \sum_{n,m} d_{nm} \pi_{nm}
    \;,
\end{align}
where $\Phi(p, q)$ represents a set of joint probability distributions $\pi_{nm}$ that adhere to the constraints $\sum_{m} \pi_{nm} = p _n$ and $\sum_{n} \pi_{nm} =q _m$. From the Kantorovich-Rubinstein duality~\cite{Villani2009,Santambrogio2015}, we can recast the Wasserstein-1 distance in a dual formulation as
\begin{align} 
    \mathcal{W}_1 (p, q) = \sup_{h \in \text{Lip}_1} \sum_{n} h_n (p _n - q _n)
    \;,
\end{align}
where $\text{Lip}_1$ denotes the set of Lipschitz functions on the graph $\mathcal{G}(V,E)$ defined by 
\begin{align} \label{eqA:definition_Lip1}
    \text{Lip}_1 \equiv \left\{ h_n | d_{nm}\ge |h_n - h_m|  \,\, \text{for all }n,m \in V \right\}
    \;.
\end{align}

With this knowledge, we aim to prove Eq.~\eqref{eq:Lip_1_other_expression}, which states that the above is equivalent to
\begin{align} \label{eqA:redefinition_Lip1} 
    \text{Lip}_1 = \left\{ h_n | 1 \ge |h_n - h_m| \,\, \text{for all }(n,m)\in E \right\}
    \;.
\end{align}
To facilitate our discussion, we introduce the notations
\begin{align} 
    &L \equiv \left\{ h_n |  d_{nm} \ge |h_n - h_m|  \,\, \text{for all }n,m \in V \right\} \;,
    \nonumber\\ &L' \equiv \left\{ h_n | 1\ge |h_n - h_m| \,\, \text{for all }(n,m)\in E \right\}
    \;.
\end{align}
If $h \in L'$, then $|h_x - h_y| \le 1$ for all $(x,y) \in E$. Therefore,  
\begin{align} \label{eqA:Lip_derivation}
    |h_n - h_m| &\le |h_n - h_{x_1}| + |h_{x_1} - h_{x_2}| + \cdots + |h_{x_k} - h_m|
    \nonumber\\ &\le \text{len}(P)
    \;,
\end{align}
where $P=(n,x_1,\cdots,x_k,m)$ denotes a path connecting $n$ and $m$. Minimizing the path length over all paths connecting $n$ and $m$ in Eq.~\eqref{eqA:Lip_derivation} leads to the inequality $|h_n - h_m|\le d_{nm}$, which imples $h \in L$. This proves $L' \subseteq L$. Conversely, if $h \in L$, it is straightforward to see that $|h_x - h_y| \le 1$ for all $(x,y) \in E$. Thus $h \in L'$, proving $L \subseteq L'$. Consequently, $L = L'$, proving Eq.~\eqref{eq:Lip_1_other_expression}.

\section{Effect of $\mathcal{B}$ on TUR tightness} \label{secA:initial_dist}

Throughout our discussions, we selected an initial distribution independent of both $s$ and $\omega$. For Langevin systems, this means fixing $\mathcal{B}=0$ in Eq.~\eqref{eq:FI_at_0}. Nevertheless, it is in principle possible to make the initial distribution depend on $s$ or $\omega$, which allows nonzero $\mathcal{B}$. Following this scheme, the XTUR~\eqref{eq:XTUR1} is modified to
\begin{align} \label{eq:modified_XTUR}
    \frac{\Omega_\tau ^2}{\text{Var}[\Theta_\tau^{\Lambda,g}]} \le \frac{ \Sigma_\tau^{\rm tot}+\mathcal{B}}{2}
    \;.
\end{align}
Setting $g = 0$, thus $\Theta_\tau^{\Lambda,g} = \mathcal{J}^\Lambda_\tau$, yields the modified version of the underdamped TUR~\eqref{eq:underdampedTUR}
\begin{align} \label{eq:modified_underdampedTUR}
    \frac{\left[(t\partial_t - s\partial_s - \omega \partial_\omega)\langle\mathcal{J}_\tau^\Lambda\rangle\right]^2}{\text{Var}[\mathcal{J}_\tau^\Lambda]} \le \frac{\Sigma^{\text{tot}}_\tau+\mathcal{B}}{2}
    \;.
\end{align}
Then, it is natural to ask how the presence or absence of $\mathcal{B}$ affects the tightness of the bound set by the TUR. To put it more precisely, we define
\begin{align} \label{eq:Qa_def}
    \mathcal{Q}_{a} \equiv \frac{\text{Var}(\mathcal{J}^\Lambda_\tau)}{\left[(t\partial_t - s\partial_s - \omega \partial_\omega)\langle\mathcal{J}_\tau^\Lambda\rangle\right]^2} (\Sigma_\tau^{\rm tot}  +\mathcal{B})
\end{align}
for the initial condition that depends on $s$ or $\omega$, and
\begin{align} \label{eq:Qb_def}
    \mathcal{Q}_{b} \equiv \frac{\text{Var}(\mathcal{J}^\Lambda_\tau)}{\left[(t\partial_t - s\partial_s - \omega \partial_\omega)\langle\mathcal{J}_\tau^\Lambda\rangle\right]^2} \Sigma_\tau^{\rm tot}
\end{align}
for the initial condition that does not depend on $s$ and $\omega$. According to the TUR, these two $\mathcal{Q}$ factors both satisfy $\mathcal{Q}_{a,b} \ge 2$. Thus, our question boils down to which $\mathcal{Q}$ factor comes closer to the lower bound $2$.

To explore this issue on a concrete basis, we set up an analytically tractable model. More specifically, we consider an underdamped Brownian particle on a one-dimensional ring subjected to a constant driving force, as described by the Langevin equation
\begin{align}
    \dot{x} &= s \frac{p}{m}\,, ~~~~
 \dot{p} = sf - \frac{\gamma}{m} p + \xi
    \;,
\end{align}
where $\langle \xi(t) \rangle=0$ and $\langle \xi(t) \xi(t') \rangle = 2T \delta(t-t')$. We focus on the steady state, so that the initial distribution must be given in the form
\begin{align}
    P^{\text{ss}} (p;s) = \frac{1}{\sqrt{2\pi mT}} \exp\left[{- \frac{1}{2mT} \left(p - \frac{smf}{\gamma} \right)^2}\right]
    \;.
\end{align}
Now, let us specify two initial distributions that lead to different forms of the TUR. The first is the $s$-dependent initial distribution $P_a \equiv P^{\text{ss}}(p;s)$, and the second is the $s$-independent counterpart $P_b \equiv P^{\text{ss}}(p;s=1)$. Given these, let us denote by $\langle\cdots\rangle_{a}$ and $\langle\cdots\rangle_{b}$ the ensemble averages with respect to the initial distributions $P_a$ and $P_b$, respectively. Then, taking $s = 1$ in the end for fair comparison, the $\mathcal{Q}$ factors are given by
\begin{align} \label{eq:QaQb}
    \mathcal{Q}_{a,b} \equiv \left.\frac{\text{Var}(\mathcal{J}_\tau^\Lambda)}{\left[(\tau \partial_\tau - s \partial_s) \langle \mathcal{J}_\tau^\Lambda \rangle_{a,b}\right]^2} \left(\Sigma_\tau^{\rm tot}  + \mathcal{B}_{a,b}\right)\right|_{s=1}
    \;,
\end{align}
where
\begin{align} \label{eq:BaBb}
    \mathcal{B}_a &= 2\left\langle \frac{f^2}{\gamma^2 T^2} \left(p - \frac{smf}{\gamma}\right)^2 \right\rangle_a = \frac{2s^2 mf^2}{\gamma^2 T}\;,\nonumber\\
    \mathcal{B}_b &= 0\;.
\end{align}


To proceed further, we choose the particle displacement $\mathcal{J}_\tau^1 = \int_0 ^\tau dt\, \dot{x} = \int_0 ^\tau dt\, s \frac{p}{m} = x(\tau) - x(0)$ as the current-like observable of interest. Using the formal solution for the particle's momentum
\begin{align}
    p(t) &= \frac{smf}{\gamma} + \left[p(0) - \frac{smf}{\gamma} \right] e^{-\frac{\gamma}{m}t}  
    \nonumber\\ &\qquad+ \sqrt{2\gamma T} \int_0 ^t dt'\,e^{-\frac{\gamma}{m} (t-t')} \xi
    \;,
\end{align}
the mean of the observable for each initial distribution is computed as
\begin{align}
    &\langle \mathcal{J}_\tau^1 \rangle_a = \frac{s^2 \tau f }{\gamma}\,,
    \nonumber\\ &\langle \mathcal{J}_\tau^1 \rangle_b = \frac{s^2 \tau f }{\gamma}+s(1-s) \left(1 - e^{-\gamma \tau /m} \right) \frac{fm}{\gamma^2}
    \;,
\end{align}
from which follow
\begin{align} \label{eqA:TUR_denominator}
    &(\tau \partial_\tau - s \partial_s) \langle \mathcal{J}_\tau^1 \rangle_a \big|_{s=1} = - \frac{\tau f}{\gamma}\,,
    \nonumber \\ &(\tau \partial_\tau - s \partial_s) \langle \mathcal{J}_\tau^1 \rangle_b \big|_{s=1} = - \frac{\tau f}{\gamma} + \left(1 - e^{-\gamma \tau /m} \right) \frac{fm}{\gamma^2}
    \;.
\end{align}

Meanwhile, as detailed in Ref.~\cite{Lee2021}, we have
\begin{align} \label{eq:sigmatauvarJ}
	\left.\Sigma_\tau ^{\text{tot}}\right|_{s=1} &= \tau f^2 / T\gamma, \nonumber\\
\left.\text{Var} (\mathcal{J}^1_\tau)\right|_{s=1} &= \frac{2\tau T}{\gamma} \left[1 - \frac{m}{\tau \gamma} \left(1 - e^{-\gamma \tau /m} \right) \right]	\;.
\end{align}

\begin{figure}[t]
\includegraphics[width=\columnwidth]{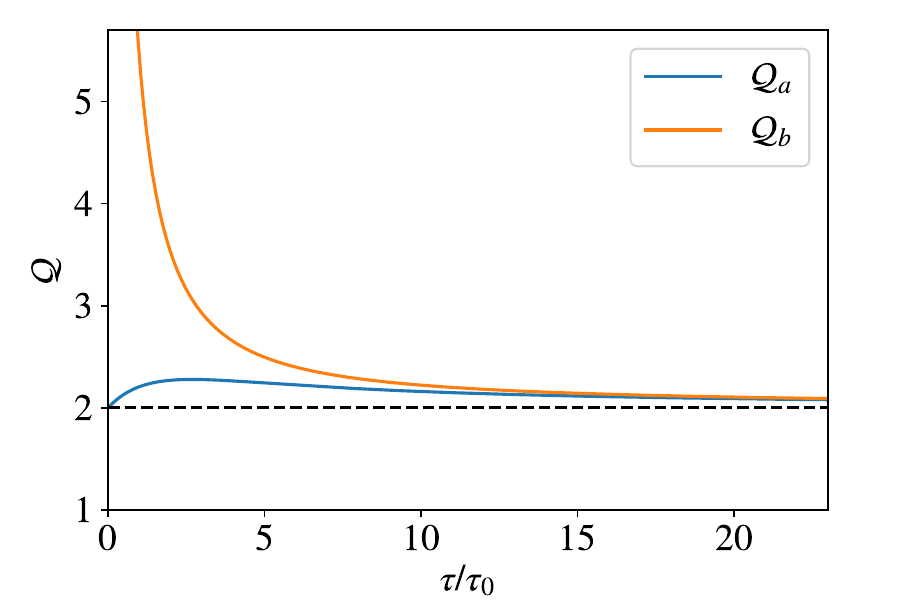}
    \caption{\label{fig:figS1} Behaviors of $\mathcal{Q}_a$ and $\mathcal{Q}_b$, obtained in Eq.~\eqref{eq:q-factors_for_ex}, as the observation time $\tau$ is varied. The dashed line at $\mathcal{Q} = 2$ corresponds to the TUR saturation.}
\end{figure}

Using Eqs.~\eqref{eq:BaBb}--\eqref{eq:sigmatauvarJ} in Eq.~\eqref{eq:QaQb}, the $\mathcal{Q}$ factors are obtained as 
\begin{align} \label{eq:q-factors_for_ex}
    &\mathcal{Q}_a = 2\left(1 + \frac{2 \tau_0}{\tau}\right) \left[1 - \frac{\tau_0}{\tau} (1 - e^{-\tau/\tau_0}) \right]\,,
    \nonumber\\ &\mathcal{Q}_b = \frac{2}{1 - \frac{\tau_0}{\tau} (1 - e^{-\tau/\tau_0})}
    \;,
\end{align}
where $\tau_0 \equiv \gamma / m$. With these, it is straightforward to show $\mathcal{Q}_b \ge \mathcal{Q}_a \ge 2$. These inequalities are verified in Fig.~\ref{fig:figS1}, which plots the $\mathcal{Q}$ factors as the observation time $\tau$ is varied. In the long-time regime, both $\mathcal{Q}$ factors approach each other and converge to $2$. In the short-time regime, however, their behaviors differ significantly: while $Q_b$ diverges to infinity, $Q_a$ converges again to $2$, thus saturating the TUR. This demonstrates that the $s$-dependent initial distribution results in a tighter bound compared to the $s$-independent initial distribution.

Beyond this specific example, it is difficult to prove in general whether the same inequalities hold for the general Langevin systems. But at least, when the observation time is short ($\tau \ll 1$), we can show that $\mathcal{Q}_b \ge \mathcal{Q}_a \ge 2$ holds for more general cases involving current-like observables of the form $\mathcal{J}_\tau ^\Lambda (\Gamma) = \int_0 ^\tau dt \,\Lambda(x,p) \circ \dot{x}$. Towards this end, we first compute
\begin{align} \label{eqA:general_observable}
    \langle \mathcal{J}_\tau ^\Lambda \rangle = \int_0 ^\tau dt \int dx dp \, \frac{s}{m} p \Lambda (x,p) P(x,p,t;s,\omega)
    \;.
\end{align}
When $\tau \ll 1$, we can use the expansion
\begin{align} \label{eqA:B_effect_expansion}
    &(\tau \partial_\tau - s \partial_s - \omega \partial_\omega)\langle \mathcal{J}_\tau^\Lambda \rangle
    \nonumber \\
    &\quad = \partial_\tau [(\tau \partial_\tau - s \partial_s - \omega \partial_\omega)\langle \mathcal{J}_\tau^\Lambda \rangle] \big|_{\tau = 0} \tau + O(\tau ^2)
    \;.
\end{align}
Using these, the denominator of the $\mathcal{Q}$ factor satisfies
\begin{align} \label{eqA:B_effect_derivation}
    &\partial_\tau [(\tau \partial_\tau - s \partial_s - \omega \partial_\omega)\langle \mathcal{J}_\tau^\Lambda \rangle] \big|_{\tau = 0} 
    \nonumber \\&\quad= - \int dx dp \,\Lambda(x,p) \,s(s\partial_s + \omega \partial_\omega) P(x,p,0;s,\omega)
    \;.
\end{align}
This vanishes when the initial distribution is independent of $s$ and $\omega$; otherwise, it is generally finite. This indicates that the denominator of the $\mathcal{Q}$ factor is $O(\tau^4)$ in the former case and $O(\tau^2)$ in the latter. The other observables in the $\mathcal{Q}$ factor typically scale as $\text{Var}(\mathcal{J}_\tau^\Lambda) \sim \tau^2$, $\Sigma_\tau ^{\text{tot}} \sim \tau$, and $\mathcal{B} \sim \tau^0$ for $\tau \ll 1$. Combining these scaling behaviors, we obtain $\mathcal{Q}_a \sim \tau^0$ for nonzero $\mathcal{B}$ and $\mathcal{Q}_b \sim \tau^{-1}$ for vanishing $\mathcal{B}$. Thus, when $\tau \ll 1$, we conclude that the TUR with nonzero $\mathcal{B}$ is tighter than the TUR with vanishing $\mathcal{B}$.

\section{Equality condition of the CSL} \label{Appendix_c}

Here, we discuss when the CSL for Langevin systems, Eq.~\eqref{eq:CSL_Wasserstein}, saturates. From Eq.~\eqref{eq:EB_to_CSL}, when $\mathcal{H} = {\rm Lip}_1$, the CSL reads
\begin{align} \label{eqA:EB_to_CSL}
    \mathcal{W}_1^2(p(0),p(\tau)) \le \Sigma_{\tau} ^{\text{tot}} \chi_\tau^{\rm Lip_1} 
    \;,
\end{align}
where $ \mathcal{W}_1 (p(0),p(\tau))= \sup_{h\in {\rm Lip_1}} |\langle h \rangle_{p(\tau)} - \langle h \rangle_{p(0)}|$ and $\chi_\tau ^{\rm Lip_1} =\sup_{h\in {\rm Lip_1}} \int_0 ^\tau dt  \,\int d \mathbf{x} (\nabla h ^{\textsf T} \mathbb{D} \nabla h )p$. We denote by $h_\mathcal{W}^*$ and $h_\chi^*$ the functions $h\in {\rm Lip_1}$ that achieve the supremum conditions for $\mathcal{W}_1$ and $\chi_\tau ^{\rm Lip_1}$, respectively. Then, we can reexpress $\mathcal{W}_1$ and $\chi_\tau ^{\rm Lip_1}$ as 
\begin{align} \label{eqA:h*}
&\mathcal{W}_1(p(0),p(\tau))= |\langle h_\mathcal{W}^* \rangle_{p(\tau)} - \langle h_\mathcal{W}^* \rangle_{p(0)}| \;, \nonumber \\
&\chi_\tau ^{\rm Lip_1} =\int_0 ^\tau dt  \int d \mathbf{x} \,({\nabla h_\chi^*}^{\textsf T} \mathbb{D} \nabla h_\chi^* )p \;.   
\end{align}
In general, $h_\mathcal{W}^*$ and $h_\chi^*$ do not need to be the same.

The equality condition of the CSL can be deduced from that of the EB since the CSL is derived from the EB. It is straightforward to check that the EB~\eqref{eq:EB} is satisfied as an equality when $\boldsymbol \Lambda =  \frac{\mathbb{D}^{-1} \mathbf{J}}{c p}$~\cite{Lee2023}, where $c$ is an arbitrary nonzero real constant. Consequently, the CSL~\eqref{eqA:EB_to_CSL} becomes an equality when
\begin{align} \label{eqA:equal_condition_CSL1}
    \nabla h_\mathcal{W}^* = \nabla h_\chi^*  = \frac{\mathbb{D}^{-1} \mathbf{J}}{c p} \equiv \nabla h^{\rm eq } 
    \;.
\end{align}

We can proceed further by focusing on the one-dimensional systems. In those cases, $\mathcal{W}_1$ is given by~\cite{Santambrogio2015}
\begin{align} \label{eqA:W_cumul}
    \mathcal{W}_1 (p(0),p(\tau)) &= \int_0^1 dy\, |F^{-1}_\tau (y) - F^{-1}_0 (y)| \nonumber \\
    &= \int_{-\infty}^\infty dx\, |F_\tau (x) - F_0 (x)|
    \;,
\end{align}
where $F_t^{-1}(y)$ is the inverse function of the cumulative distribution function $F_t(x) \equiv \int_{-\infty}^x dx^\prime \,p(x^\prime,t)$. From Eq.~\eqref{eqA:h*}, we also have
\begin{align} \label{eqA:W_another_expression}
    \mathcal{W}_1 (p(0),p(\tau)) 
    & = \left| \int_{-\infty}^\infty dx \, h_\mathcal{W}^* [p(x,\tau) - p(x,0)] \right| 
     \nonumber \\
    & = \left| \int_{-\infty}^\infty dx \,(\partial_x h_\mathcal{W}^*)[F_\tau(x) - F_0(x)] \right|,
\end{align}
where the second equality follows from integration by parts. By equating Eqs.~\eqref{eqA:W_cumul} and \eqref{eqA:W_another_expression}, we find
\begin{align}
    \partial_x h_\mathcal{W}^* = \text{sgn}[F_0(x) - F_\tau(x)] 
    \;,
\end{align}
which implies $\partial_x h_\mathcal{W}^* = \pm 1$. 

For the one-dimensional systems, one can easily determine $\partial_x h_\chi^*$ satisfying Eq.~\eqref{eqA:h*}. In this case, the diffusion matrix $\mathbb{D}$ can be expressed as a scalar constant $D$, and every function $h\in {\rm Lip_1}$ satisfies $|\partial_x h| \leq 1 $. Then, the supremum condition for $\chi_\tau^{\rm Lip_1}$ is achieved when $\partial_x h_\chi^* = \pm 1$. Combining this with Eq.~\eqref{eqA:equal_condition_CSL1}, the equality condition of the CSL for the one-dimensional Langevin systems is found as
\begin{align} \label{eqA:equal_condition_CSL2}
    \partial_x h_\mathcal{W}^* = \partial_x h_\chi^* =  \frac{J}{c D  p} = \partial_x h^{\rm eq} = \pm 1 \;.
\end{align}

Using the above equation, we can derive an external protocol that saturates the CSL. We focus on the case where $\partial_x h^{\rm eq} = 1$. For the inhomogeneous temperature field $D= T(x)/\gamma$, the particle distribution follows the Fokker-Planck equation
\begin{align} \label{eqA:equalityFP}
    \partial_t p(x,t) = -\partial_x J =  -c^\prime \partial_x [T(x) p(x,t)] \;,
\end{align}
where $c^\prime \equiv c/\gamma$. Separating the variables as $p(x,t) = X(x) \mathcal{T}(t)$, we obtain
\begin{align} \label{eqA:sv_position}
    -\frac{1}{c^\prime \mathcal{T}} \frac{d\mathcal{T}}{dt} = \frac{1}{X}\frac{d(T X)}{dx} = i \omega
    \;,
\end{align}
where $\omega$ is a constant independent of space and time. This equation is solved by the Ans\"{a}tze
\begin{align}
    \mathcal{T}(t)=e^{-i c^\prime \omega t},\,\,X(x)=\frac{1}{T(x)} \exp\left(i\omega \int_0 ^x dx' \frac{1}{T(x')} \right)
    \;,
\end{align}
whose linear combinations yield the general solution
\begin{align} \label{eqA:sol_p1}
    p(x,t) = \frac{1}{T(x)} \int d\omega \,A(\omega) \exp\!\left[i\omega \int_0 ^x dx' \frac{1}{T(x')} - i c^\prime \omega t \right].
\end{align}
Setting $t = 0$, the coefficient $A(\omega)$ is determined via the inverse Fourier transform
\begin{align} \label{eqA:A}
    A(\omega) = \frac{1}{2\pi} \int dy T(g^{-1}(y)) p_0 (g^{-1}(y)) e^{-i\omega y}
    \;,
\end{align}
where we have defined $p_0 (x) \equiv p(x, 0)$, $g(x) \equiv \int_0 ^x dx' \, 1/T(x')$, and its inverse $g^{-1}(y)$. Plugging Eq.~\eqref{eqA:A} into Eq.~\eqref{eqA:sol_p1} leads to
\begin{align} \label{eqA:sol_p2}
    p(x,t) = \frac{1}{T(x)} T(g^{-1}(g(x) - c^\prime t)) p_0 (g^{-1}(g(x) - c^\prime t)) \;.
\end{align}

If the protocol $V_{\rm EQ} (x,t)$ makes $p(x,t)$ evolve according to the above equation, then the CSL holds as an equality. Combining the equation with the relation
\begin{align}
    J = c^\prime T p = -(\partial_x V_{\rm EQ}) p - \partial_x (T p) \;,
\end{align}
we finally obtain
\begin{align} \label{eqA:V_EQ}
    V_{\text{EQ}} (x,t) = - \int_{x_0} ^x dx' \left[ c^\prime T(x') + \frac{\partial_x [T(x')p(x',t)]}{p(x',t)} \right]
    \;,
\end{align}
given that $x_0$ is a constant marking the boundary at which the potential stays zero.



\end{appendix}

\bibliography{unified_hierarchy}

\end{document}